\documentclass[%
    twocolumn,
    superscriptaddress,
    bitnotes,
    amsmath, amssymb, 
    aps, 
    prx,
]{revtex4-2}

\usepackage{changes}
\usepackage{graphicx}       
\usepackage{dcolumn}        
\usepackage{bm}             
\usepackage{color}
\usepackage{subfigure}
\usepackage{appendix}
\usepackage[utf8]{inputenc}
\usepackage{hyperref}
\hypersetup{
    colorlinks=true,
    linkcolor=blue,
    filecolor=magenta,      
    urlcolor=cyan,
    pdfpagemode=FullScreen,
}

\newcommand\COMMENTED[1] {}

\begin{document}
\title{Temperature Dependence of Spin and Charge Orders\\ in the Doped Two-Dimensional Hubbard Model}

\author{Bo Xiao}
\email{bxiao@flatironinstitute.org}
\affiliation{Center for Computational Quantum Physics, Flatiron Institute, 162 Fifth Avenue, New York, New York, 10010, USA}

\author{Yuan-Yao He}
\email{heyuanyao@nwu.edu.cn}
\affiliation{Institute of Modern Physics, Northwest University, Xi'an 710127, China}
\affiliation{Peng Huanwu Center for Fundamental Theory, Xian 710127, China}
\affiliation{Center for Computational Quantum Physics, Flatiron Institute, 162 Fifth Avenue, New York, New York, 10010, USA}

\author{Antoine Georges}
\email{ageorges@flatironinstitute.org}
\affiliation{Center for Computational Quantum Physics, Flatiron Institute, 162 Fifth Avenue, New York, New York, 10010, USA}
\affiliation{Coll\`{e}ge de France, 11 place Marcelin Berthelot, 75005 Paris, France}
\affiliation{CPHT, CNRS, \'{E}cole Polytechnique, IP Paris, F-91128 Palaiseau, France}
\affiliation{DQMP, Universit\'{e} de Gen\`{e}ve, 24 quai Ernest Ansermet, CH-1211 Gen\`{e}ve, Switzerland}

\author{Shiwei Zhang}
\email{szhang@flatironinstitute.org}
\affiliation{Center for Computational Quantum Physics, Flatiron Institute, 162 Fifth Avenue, New York, New York, 10010, USA}

\date{\today}

\begin{abstract}
    Competing and intertwined orders including inhomogeneous patterns of spin and charge are observed in many correlated electron materials, such as  high-temperature superconductors.  
    Introducing a new development of the constrained-path auxiliary-field quantum Monte Carlo (AFQMC) method, we study the interplay between thermal and quantum fluctuations in the two-dimensional Hubbard model.  We obtain an accurate and systematic characterization of 
    the evolution of the spin and charge correlations as a function of temperature $T$ and how it connects to the ground state, at three representative hole doping levels $\delta = 1/5$, $1/8$, and $1/10$.  We find increasing short-range commensurate antiferromagnetic correlations as $T$ is lowered.  As the correlation length grows sufficiently large, a modulated spin-density-wave (SDW) appears.  
    At $\delta = 1/5$ and $U/t = 6$, the SDW saturates and remains short-ranged as $T \rightarrow 0$.  In contrast, at $\delta = 1/8$, $1/10$ and $U/t = 8$ this evolves into a ground-state stripe phase. We study the relation between spin and charge orders and find that formation of charge order appears to be driven by that of the spin order.  We identify a finite-temperature phase transition below which charge ordering sets in and discuss the implications of our results for the nature of this transition. 
\end{abstract}

\maketitle
\section{\label{sec:intro} INTRODUCTION}

The complex interplay of spin, charge and pairing correlations is a common feature of many strongly correlated materials, from manganites to cuprates~\cite{Dagotto2005, Orenstein2000, Lee2006, Imada1998, Scalapino2012}.  In the latter case, inhomogeneous orders such as stripes 
are a central concept~\cite{Kivelson2003}. Spin and charge, rather than being uniformly distributed, form intricate collective ordering patterns which often have characteristic wavelengths spanning multiple lattice spacings. For example, stripe orders exhibit antiferromagnetic (AFM) regions separated by $\pi$-phase shifts that reverse the sublattice occupations across lines where the holes accumulate.  
Related spin-density wave (SDW) states also show modulated AFM correlation, but with more uniform charge distribution in which holes spread into regions away from the spin phase shift lines.  These states tend to be collinear in the spin order, and 
the wavelengths of the spin and charge modulation are 
related:  $ \lambda_{\rm spin}=2\lambda_{\rm charge}$~\cite{Emery1999, Comin2016}.  Such states are the outcome of a
compromise and balance among the AFM interactions, Coulomb interactions and kinetic energies \cite{Zaanen2000}.

Some families of cuprates such as Nd- or Ba- substituted $\rm La_{2-x}Sr_{x}CuO_{4}$ (LSCO) indeed display long-range stripe ordering~\cite{Tranquada1995, Tranquada1997, Fujita2002, Kivelson2003, Zaanen1999}.  Taking a broader perspective, charge ordering is a commonly observed feature of the low-doping ground-state of cuprate superconductors when a magnetic field is applied to suppress superconductivity~\cite{Sachdev2002,Wu2011,Comin2016}.

The one-band Hubbard model and related models have played a key role in the study of stripe and SDW 
orders~\cite{Arovas2021, Qin2021,Georges2021}. Indications of these orders were first seen in Hartree-Fock calculations in the two-dimensional (2D) Hubbard model~\cite{Zaanen1989, Poilblanc1989, Machida1989, Schulz1990, Kato1990}. Calculations in the related $t$-$J$ model with a hole doping $\delta = 1 / 8$ using the density matrix renormalization group (DMRG) showed evidence for 
$\pi$ phase-shifted AFM regions separated by domain walls~\cite{White1998}. Further DMRG studies in the doped Hubbard model on cylinders found the existence of a stripe state~\cite{White2003}. Ground-state constrained-path (CP) auxiliary-field quantum Monte Carlo (AFQMC) calculations in the Hubbard model found inhomogeneous spin and charge orders~\cite{Chang2008}, which were shown to be SDW states at $U/t=4$ and stripe states at larger $U$~\cite{Chang2010}. At the thermodynamic limit, these states were shown~\cite{Chang2010} to have modulation along the $x$- or $y$- direction, with wavelength of $2/\delta$ for the spin and $1/\delta$ for the charge. A recent multi-method study~\cite{Zheng2017} using four different computational methods confirmed that the ground state of the 2D Hubbard model at $\delta = 1/8$ doping and strong $U$ is indeed a vertical ($x$ or $y$) stripe state with spin modulation wavelength of $16$ and charge modulation wavelength of $8$, i.e., {\it filled\/} stripes. Note that this stripe pattern is somewhat different from the one observed in real materials~\cite{Tranquada1995, Tranquada1997, Kivelson2003, Raczkowski2010, Jiang2020}.

Determining and understanding the full phase diagram of the Hubbard model has remained an outstanding challenge. Computational methods need to reach high accuracy and sufficiently large system sizes to accommodate collective states whose modulating wavelengths tend to be large.  Ground-state properties sensitively depend on the delicate balance of small energy differences. Often the signal is comparable to or smaller than the uncertainties in the numerical methods (due, e.g., to finite supercell size effects, finite bond dimensions, limitation on the temperatures and statistical error bars, approximations in the computational scheme, etc.).  Recent years have seen major advances in computational methods which have led to considerable progress in the determination of the ground-state of the Hubbard model. For example, studies by DMRG~\cite{Ehlers2017, Jiang2020}, infinite project-entangled pair states (iPEPS)~\cite{Corboz2014}, density matrix embedding theory (DMET)~\cite{Zheng2016}, inhomogeneous (unrestricted) dynamical mean field theory (DMFT)~\cite{Peters2014} and its cluster extensions~\cite{Vanhala_2018}, variational Monte Carlo~\cite{Ido2018, Becca2019, Sorella2021}, AFQMC~\cite{Xu2022} etc.~have all indicated that stripe or modulated spin orders feature prominently throughout a large portion of the ground state phase diagram in the 2D Hubbard model, either as the true ground state or energetically very close by. More subtle and controversial has been the existence of superconducting order in the model~\cite{Maier2005_PRL, Maier2005, Gull2012, Gull2013, Sordi2012, Sorella2021}. It is clear that a strong interplay exists between superconductivity and the stripe and SDW orders, and recent results indicate that there is a competition which appears to favor the latter in the ground state of the pure 2D 
Hubbard model --- namely in its original form without hopping beyond nearest neighbors --- at intermediate to strong coupling and near optimal doping~\cite{Qin2020}.

Understanding the temperature dependence of the magnetic and charge orders is of fundamental importance.  Information about the temperature evolution of stripe and SDW correlations is crucial for making connections with experimental observations. At finite temperatures, new fascinating phenomena arise such as the pseudogap regime whose understanding requires reliable characterization of spin and charge properties.  However, the computational challenges are outstanding. In addition to all the requirements for reliable and predictive calculations at $T=0\,$K, we must now not only retain thermal fluctuations but reach sufficiently low temperatures.  As a result, much less is known about the properties of the magnetic and, especially charge correlations in the doped Hubbard model at finite temperatures.

Recently there has been a flurry of activities to study the finite-temperature properties of stripes as well as 
spin and charge correlations in the doped Hubbard model, including calculations using determinant Quantum Monte Carlo (DQMC)~\cite{Huang2017, Huang2018}, minimally entangled typical thermal state (METTS)~\cite{Wietek2021}, the dynamical cluster approximation (DCA)~\cite{Mai2022}, and connected determinant  diagrammatic Monte Carlo (CDet)~\cite{Simkovic2021}.  This has created an exciting synergy, from which a consistent picture is emerging on the temperature evolution of the antiferromagnetic spin correlations and charge inhomogeneities in the doped Hubbard model. The picture is far from complete, however. Despite impressive advances in the computational methods, technical hurdles in each study limited the scope of questions that could be addressed. These hurdles included the sizes of the simulation cells or clusters that could be accessed (which limit the detection of long-range correlations), the use of relatively narrow (width-4) cylindrical cells in some studies (whose quasi-one-dimensional nature may not represent the physical behavior in two-dimensions), restriction to relatively high temperatures, etc.  Important questions remain on how the spin and charge correlations connect to the ground state, whether and how the system develops into long-range-ordered state, and the interplay and connection between spin and charge ordering as a function of temperature.

In this paper, we address these questions by employing finite-temperature AFQMC \cite{Zhang1999, Pavarini2019} with a self-consistent constraint~\cite{He2019} and by formulating a more flexible and powerful self-consistency scheme at finite temperatures which introduces an effective temperature in the constraint.  Our new approach allows for accurate computations with much larger simulation cells at much lower temperatures than previously possible.  We employ two different ways to probe the spin and charge correlations, and combine them with a careful investigation of the size dependence to extract the properties in the two-dimensional thermodynamic limit. We focus on the pure Hubbard model ($t'=0$) and investigate three representative doping levels in the parameter regime most relevant to cuprates: $\delta = 1 / 5$ ($U/t = 6$), $\delta = 1 / 8$ and $1 / 10$ ($U/t = 8$).  In each case we systematically determine the temperature dependence and how the system evolves into its corresponding ground state.  We find that the $\delta = 1 / 5$ case shows qualitatively and fundamentally different behavior from the other two, which develop long-range ground-state order.
Notably, we answer a long-standing open question in the field by 
identifying the onset temperature at which long-range charge order forms. We discuss possible scenarios regarding the interplay and nature of the charge- and spin-ordering transitions.

The paper is organized as follow. In Sec.~\ref{sec:methods}, we introduce the Hubbard Hamiltonian and give a brief overview of our computational approach, followed by a discussion and illustration of the new finite-temperature approach that optimizes the trial density matrix self-consistently using an effective temperature.  Our results are presented and discussed in Sec.~\ref{sec:results}.
In Sec.~\ref{ssec:spin_correlation}, we examine the temperature dependence of the spin order in real space.  This is complemented by momentum space information in Sec.~\ref{ssec:spin_susceptibility}.  In Sec.~\ref{ssec:short_range_order}, we focus on the overdoped regime at $\delta = 1 / 5$ and $U = 6$, showing the temperature evolution of a system which has only short-range modulated spin correlations in the ground state.
Our findings of the interplay between onsets of spin and charge ordering and 
the critical $T_{c}$ for charge long-range order at $\delta = 1 / 8$ and $U = 8$ 
are then presented in Sec.~\ref{ssec:long_range_order}.
We conclude in Sec.~\ref{sec:conclusion}.

\section{\label{sec:methods} MODEL AND METHOD}

The Hubbard Hamiltonian \cite{Hubbard1963, Hubbard1964, Arovas2021, Qin2021}, which was originally introduced to explain the itinerant ferromagnetism of transition-metal monoxides, serves as a minimal microscopic model to study several key features in strongly interacting quantum systems.  It takes the form
\begin{eqnarray}
    \begin{aligned}
        H = & -t\sum_{\langle {\bf i}, {\bf j} \rangle, \sigma} (c_{{\bf i} \sigma}^{\dag}c_{{\bf j} \sigma} + c_{{\bf j} \sigma}^{\dag}c_{{\bf i} \sigma}) - \mu \sum_{{\bf i}} n_{{\bf i}} \\ 
        & + U \sum_{{\bf i}} \left(n_{{\bf i} \uparrow} - \frac{1}{2} \right) \left(n_{{\bf i} \downarrow} - \frac{1}{2} \right).    \label{Hamiltonian}      
    \end{aligned}
\end{eqnarray}
We consider a 2D lattice with $t$ denoting the nearest-neighbor hopping amplitude and $\langle {\bf i}, {\bf j} \rangle$ denoting a pair of nearest-neighbor lattice sites ${\bf i}$ and ${\bf j}$.  We choose 
$t = 1$ to set the unit of energy.  $c_{{\bf i}, \sigma}^{\dag}$ ($c_{{\bf i}, \sigma}^{}$) is the creation (annihilation) operator of electrons with spin $\sigma$ at site 
${\bf i} = (i_{x}, i_{y})$,
and $n_{{\bf i}, \sigma} = c_{{\bf i}, \sigma}^{\dag}c_{{\bf i}, \sigma}^{}$ denotes the number operator. $U$ is an on-site interaction between spin-up and spin-down electrons. The chemical potential $\mu$ tunes the electron density and $\mu = 0$ corresponds to half-filling due to the choice of the particle-hole symmetric form for the interaction.  
The hole doping is defined as $\delta=1-\rho$ 
where $\rho = N_{e} / N$ denotes the electron density, $N_{e}$ is the particle number, and $N = L_{x} \times L_{y}$ is the number of sites in the simulation cell.

We solve this model by a state-of-the-art AFQMC method which controls the fermion sign problem with a constraint on the paths in auxiliary-field space \cite{Zhang1999, Pavarini2019}.  In the standard DQMC algorithm \cite{Blankenbecler1981}, the interaction part of the Hamiltonian is formulated in terms of single-particle operators, after applying the Hubbard-Stratonovich (HS) transformation \cite{Hubbard1959, Hirsch1985} to the two-body terms and replacing them by one-body interactions with a set of auxiliary fields.  The partition function, $e^{-\beta H}$, where $\beta=1/T$ is inverse temperature, can be treated as a many-dimensional path integral over random auxiliary fields, which is computed using Monte Carlo (MC) techniques. The sign problem occurs because different configurations of the auxiliary-fields, i.e., different paths, lead to contributions to the partition function with different signs. As the temperature is decreased (i.e., the length of the path in imaginary-time increased), the average sign of the contributions approaches zero exponentially \cite{White1989, Loh1990, Schmidt1987, Tarat2022, Mondaini2022}. The sign problem fundamentally limits the accessible temperatures and lattice sizes in simulations of strongly correlated systems by standard DQMC algorithms.

The AFQMC approach we employ shares with DQMC the decoupling of the two-body interactions by HS transformation and replacing them with one-body interactions in auxiliary-fields. However, it reformulates the partition function as a path integral over a constrained portion of the paths in auxiliary-field space.  The full path-integral gives the many-body partition function via an over-complete space. There exists an exact sign or gauge condition for a reduced subset of paths which preserve the exactness of the partition function \cite{Zhang1999, Pavarini2019}.  In practice this condition is realized approximately by a trial density matrix. In the limit $T=0\,$K, this approach reduces to the ground-state CP method (or more generally, the phaseless AFQMC \cite{Zhang2003, Pavarini2019} for arbitrary two-body Hamiltonians such as those in electronic structure calculations).  There exists a large body of work which shows that this approach leads to highly accurate results in the ground state \cite{Zhang1999, LeBlanc2015, Chang2008, Shi2014, Qin2016Benchmark}. Recently a self-consistent constraint has been developed for ground-state \cite{Qin2016} and finite-$T$ \cite{He2019} calculations. By coupling the AFQMC calculation to an independent particle (IP) Hamiltonian, an optimized constraint from single-particle wave function or density matrix can be achieved, which further reduces the systematic error. In this paper we propose an improved self-consistent scheme to further enhance the accuracy of the method in the most challenging cases at finite temperature, which we discuss further in Sec.~\ref{ssec:SC-finiteT} below.

\subsection{Two different approaches to characterize spin and charge orders}
\label{ssec:2approaches}

Orders in the Hubbard model are in general quite delicate, because of competitions and the small energy scales which separate them. At finite-temperatures, the spin and charge orders have collective modes which all decay with distance and are sensitive to finite sizes and boundary conditions and which thus require exquisite accuracy to determine. In this work, we use two different approaches to characterize these orders, and compute several quantities to probe them under each approach.

We perform AFQMC calculations in 2D using either supercells
under periodic boundary condition (PBC) in both $x$ and $y$ directions or cylindrical cells which are periodic along $y$ but has open boundary condition (OBC) along the $x$-direction.  In the first case, we measure various correlation functions, for instance the equal-time  spin-spin correlation in real space,
\begin{eqnarray}
    C_{s}({\bm{\ell}}) = \frac{1}{N} \sum_{{\bf i}} \langle S_{z}({\bf i}) S_{z}({\bf i} + {\bm{\ell}}) \rangle,
\end{eqnarray}
where $\hat{S}_{z}({\bf i}) = \frac{1}{2}(n_{{\bf i} \uparrow} - n_{{\bf i} \downarrow})$ is the z-component of the spin operator at site ${\bf i}$. This type of measurements are employed in Sec.~\ref{ssec:spin_correlation}. 
Imaginary-time-dependent correlation functions can also be measured.
For example the static  spin susceptibility discussed in  Sec.~\ref{ssec:spin_susceptibility}, $\chi_{s}({\bf q})$, is 
\begin{eqnarray}
    \chi_{s} ({\bf q}) = \sum_{\bm{\ell}} e^{i {\bf q} \cdot \bm{\ell}} \int_{0}^{\beta}  \langle S_{z}({\bm{\ell}}, \tau) S_{z}({\bf 0}, 0) \rangle d\tau,
\end{eqnarray}
which is the Fourier transform of the 
unequal-time spin correlation function integrated over imaginary time.  The charge correlation functions and susceptibilities can be calculated in a similar way.

When the order to be detected is small, it is especially challenging to extract signals from correlation functions (and quantities derived from them such as structure factors and susceptibilities), because one is essentially measuring  the {\it square} of the (small) order parameter.  To ameliorate this problem, we take a complementary approach to the fully PBC calculations discussed above.  We introduce ``pinning fields" to break translational symmetry in one direction ($x$) (or in both $x$ and $y$ directions in certain cases as needed), so that we can measure a local ``order parameter" in these systems as a proxy to the two-body correlation functions in translationally-invariant systems in the large size limit. 
For example, we can apply AFM pinning fields to the left edge of the simulation cell 
by adding the following term to the original Hamiltonian in Eq.~(\ref{Hamiltonian}):
\begin{eqnarray} 
\label{eq:S-pinning}
    H_{s} = \sum_{{\bf i}, \sigma} v_{{\bf i} \sigma} n_{{\bf i} \sigma} 
\end{eqnarray}
where $v_{i_{x} \uparrow} = -v_{i_{x} \downarrow} = (-1)^{i_{y}} v$ for all sites with $i_{x} = 0$.  These spin pinning fields break the translational and $\rm SU(2)$ symmetries, allowing the direct detection of the order parameter \cite{White2007, Assaad2013};
however, care must be taken to remove any bias they introduce, as we further discuss below.

The approach of pinning fields is used to obtain the results in Sec.~\ref{ssec:short_range_order} and some of the results
in Sec.~\ref{ssec:long_range_order}.  
We can similarly add charge ``pinning fields", in the form $H_{c} =\sum_{{\bf i}} P_{{\bf i}} n_{{\bf i}}$. 
For example, we can apply a fully periodic potential to the whole simulation cell:
\begin{eqnarray}
\label{eq:C-pinning}
   H_{c} = P\sum_{i_{x}, i_{y}, \sigma} \sin \left(\kappa i_{x} + \phi \right) n_{(i_{x}, i_{y}), \sigma},
\end{eqnarray}
 where $\kappa = 2\pi / \lambda$ is the wave number corresponding to the expected wavelength of the
 charge order, in order to probe the response of the system to the perturbation.
 This particular form is used 
 to study the interplay between charge and spin orders in  Sec.~\ref{ssec:long_range_order}.
In the presence of pinning fields, we preserve symmetry in the y direction by using PBC.  
In these cases, we can average over $i_y$ the staggered spin density 
 \begin{eqnarray}
    \bar{S}_{z}(i_{x}) = \frac{1}{L_{y}} \sum_{i_{y} = 0}^{L_{y} - 1}(-1)^{i_{x} + i_{y}} \langle S_{z}(i_{x}, i_{y})\rangle, 
    \label{eqn:eqn6}
 \end{eqnarray}
and hole density 
\begin{eqnarray}
   \bar{h}(i_{x}) =1- \frac{1}{L_{y}} \sum_{i_{y} = 0}^{L_{y} - 1} \langle n_{{\bf i},\uparrow} + n_{{\bf i},\downarrow} \rangle
\end{eqnarray} 
to help improve statistics in detecting the spin and charge orders.  The effect of the edge pinning field can be minimized or removed by separate calculations
using smaller values of $v$. A robust order shows no variation with $v$ except locally 
at the edge \cite{Xu2022}; 
this can be used in combination of additional calculations with increasing system sizes, 
especially for determining the long-range behavior. 
When a full pinning field is applied to the simulation cell, such as in 
Eq.~(\ref{eq:C-pinning}),
an extrapolation must be performed to $P\rightarrow 0$ to obtain the correct response.

\subsection{Controlling the sign problem at finite temperature: self-consistent constraint}
\label{ssec:SC-finiteT}

In AFQMC 
the simplest trial density matrix we use is the non-interacting type, defined
by the following trial Hamiltonian \cite{Zhang1999,He2019}
\begin{eqnarray}
    \hspace{-12pt}
    H_{T} = -t \sum_{\langle {\bf i},{\bf j} \rangle \sigma}\left(c_{{\bf i}\sigma}^{\dag}c_{{\bf j}\sigma}^{} + c_{{\bf j} \sigma}^{\dag}c_{{\bf i}\sigma}^{} \right) 
    - \sum_{{\bf i}} \mu_{{\bf i}, T} \left(n_{{\bf i} \uparrow} + n_{{\bf i} \downarrow} \right),  \label{eqn:noninteracting}
\end{eqnarray}
where $\mu_{{\bf i}, T}$ needs to be tuned such that the electron density for this single-particle Hamiltonian is equal to that for the original many-body Hamiltonian.  In this paper, we set $\mu_{{\bf i}, T} = \mu_{T}$ for simplicity and preserve translational invariance.  This type of trial density matrix is employed in our first approach discussed above in Sec.~\ref{ssec:2approaches}, namely in fully periodic calculations.

\begin{figure}
    \centering
    \includegraphics[width = .5\textwidth]{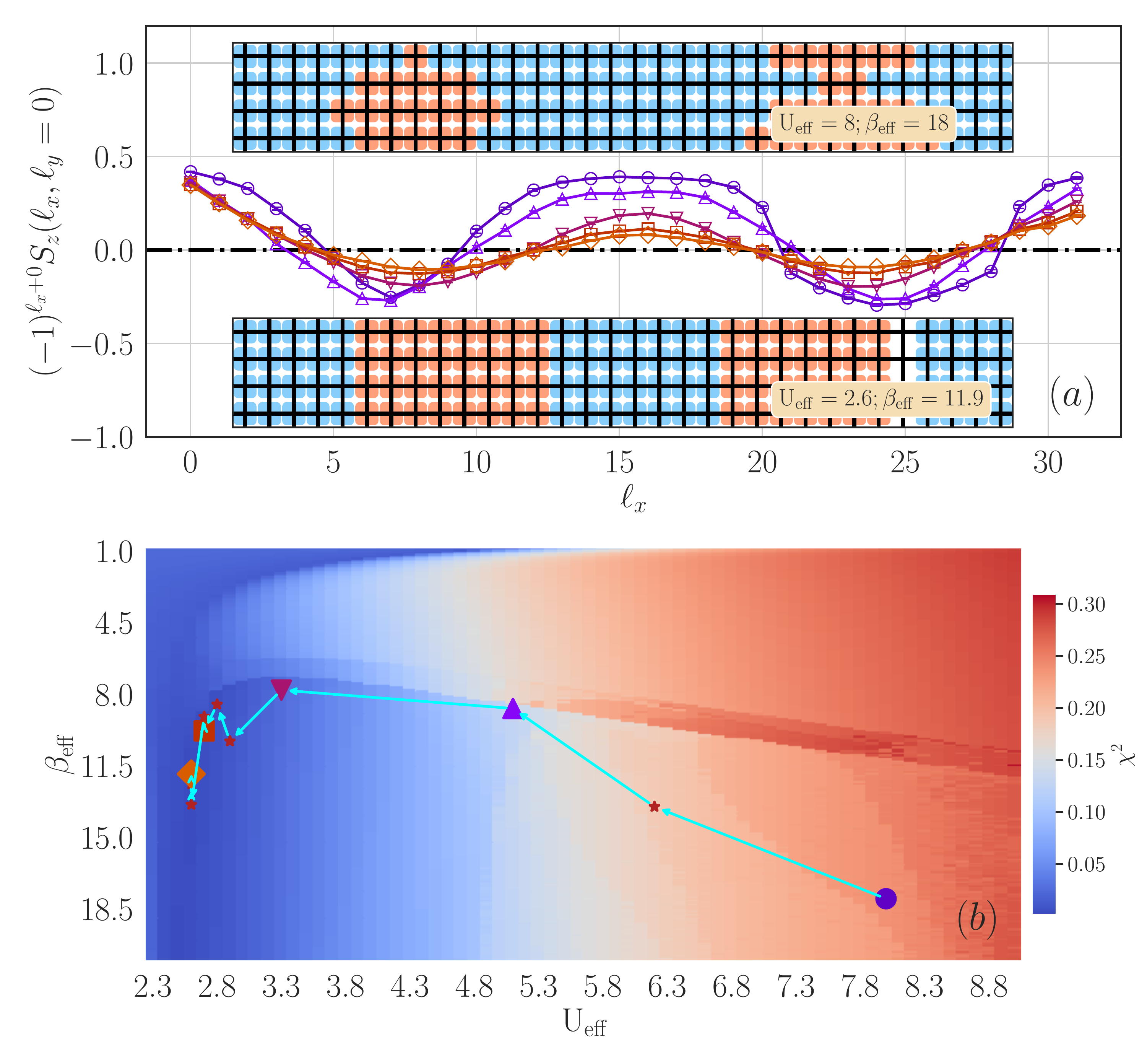}
    \caption{Systematic improvement of the AFQMC accuracy using the self-consistent constraint.  (a) The main plot shows the staggered spin density $(-1)^{\ell_{x} + 0}S_{z}(\ell_{x}, \ell_{y} = 0)$ vs.~site $\ell_{x}$, obtained from AFQMC at five representative steps in the self-consistent iteration. Statistical error bars are much smaller than symbol size.  The insets show visualization of the real-space spin density obtained from AFQMC in the first and last iterations (corresponding to the circle and diamond in the second panel, respectively) . Blue and orange colors mark  regions of AFM spin densities with a phase change between them. (White color means the magnitude of the spin density is smaller than twice the AFQMC statistical error bars.)  (b) The trajectory of how  $(U_{\rm eff}, \beta_{\rm eff})$ evolve in the self-consistent procedure is shown on the heatmap of $\chi^{2}(U_{\rm eff}, \beta_{\rm eff})$.  Each solid symbol denotes the effective interaction 
    and temperature parameters of the IP Hamiltonian at one step of the iteration, and the light blue arrows indication the direction  in the iteration. The five selected steps shown in panel (a) are indicated here by their corresponding symbol shapes, while all other steps are shown as stars.  The system is $32 \times 4$ with $\delta = 1/8$ and $U = 8$, at inverse temperature  $\beta = 18$.}
    \label{fig:fig1}
\end{figure}

To improve the CP approximation systematically, a self-consistent constraint can be employed, by coupling the AFQMC calculation with an effective independent-particle (IP) calculation \cite{Qin2016, He2019}.  
If we choose unrestricted Hartree-Fock (UHF) as the form for our IP calculation, 
the IP Hamiltonian that corresponds to the many-body one in Eq.(\ref{Hamiltonian}) 
takes the form
\begin{eqnarray}
    \begin{aligned}
    H_{\rm IP}  = & -t \sum_{\langle {\bf i}, {\bf j} \rangle, \sigma} (c_{{\bf i }\sigma}^{\dag} c_{{\bf j}\sigma}^{} 
        + c_{{\bf j} \sigma}^{\dag} c_{{\bf i} \sigma}^{}) \\
        & + \sum_{{\bf i} \sigma}\left[U_{\rm eff} \left(\langle n_{{\bf i} \bar{\sigma}} \rangle - \frac{1}{2}\right) 
          - \mu_{\rm eff} \right] n_{{\bf i} \sigma},  \label{IP}
    \end{aligned}
\end{eqnarray}
where $\bar{\sigma}$ denotes the opposite spin of $\sigma$.
The $\rm SU(2)$ spin rotational symmetry is reduced to $\rm U(1)$ symmetry because we assume the z-axis as the quantization direction.  The effective chemical potential $\mu_{\rm eff}$ tunes the particle number in the grand canonical ensemble. The effective interaction strength $U_{\rm eff}$, as well as the mean-field spin densities, $\{ \langle n_{{\bf i} \sigma} \rangle \}$, are to be determined through the self-consistent iteration with the AFQMC calculations, as discussed further below.  

The self-consistent constraint is applied in  our second approach discussed in Sec.~\ref{ssec:2approaches}, namely with simulation cells in which a symmetry-breaking pinning field is applied. The same external potential $ H_{s}$ or $ H_{c}$,  in Eq.~(\ref{eq:S-pinning}) or (\ref{eq:C-pinning}), is applied to both the many-body Hamiltonian for AFQMC and the effective Hamiltonian in the IP calculation.  With a pinning field applied at the edge, 
as mentioned earlier, the symmetry-breaking allows us to use the 
behavior of the many-body local ``order parameters" (such as the one-body spin or density) versus distance as a probe to characterize the two-body correlations in the corresponding translationally invariant system.  When a small external potential, e.g., the one in Eq.~(\ref{eq:C-pinning}), is applied to the whole simulation cell, it allows us to do linear response study of the order built into the potential.

We start the self-consistent procedure with a trial 
density matrix $e^{-\tau H_T}$, where $\tau\in (0,\beta)$ is a variable denoting 
the imaginary time where the constraint is applied \cite{Zhang1999,He2019}, and $H_T$ can be
taken as a simple mean-field Hamiltonian, e.g., the non-interacting or UHF Hamiltonian.
Using this trial density matrix as a constraint,
an AFQMC calculation is carried out 
(which always uses the physical $\beta$ and $U$ values),
and spin-resolved electron densities $\{ \langle n_{{\bf i} \sigma} \rangle_{\rm QMC} \}$ are computed.  
We then search for the best IP density matrix $e^{-\beta_{\rm eff}H_{\rm IP}(U_{\rm eff})}$ which produces spin densities
$\{ \langle n_{{\bf i} \sigma} \rangle_{\rm IP}\}$
closest to the AFQMC result. In other words, we minimize
\begin{eqnarray}
    \chi^{2}(U_{\rm eff},\beta_{\rm eff}) = \frac{1}{N} \sum_{{\bf i}, \sigma} \left(\langle n_{{\bf i} \sigma} \rangle_{\rm IP} - \langle n_{{\bf i} \sigma} \rangle_{\rm QMC} \right)^{2},  \label{eqn:self-consistent}
\end{eqnarray}
as a function of the effective interaction and effective temperature in the IP calculation. 
The IP solution
with optimal $U_{\rm eff}$ and $\beta_{\rm eff}$ 
determined from Eq.~(\ref{eqn:self-consistent}), including 
the value of $U_{\rm eff}$ and the resulting spin densities, 
is then fed as the input trial Hamiltonian to
the new AFQMC calculation.  We perform this  procedure iteratively until $(U_{\rm eff}, \beta_{\rm eff})$ have converged.

Before illustrating the self-consistent procedure with an example, we comment on two technical 
aspects. First, instead of using the IP spin densities in the trial 
Hamiltonian, we could also feed directly the AFQMC results from the
previous iteration~\cite{Qin2016}; however, the statistical error 
of the QMC results could affect their accuracy as a constraint, and 
reducing the error bars during the self-consistency would incur 
additional computational cost.The introduction of the effective inverse temperature $\beta_{\rm eff}$ allows the IP calculation to reproduce the QMC densities better. Second, 
we find it convenient and efficient to 
 perform extensive UHF calculations prior to the self-consistent procedure, to set up a "look-up" table of the spin densities on
 a two-dimensional grid of $(U_{\rm eff}, \beta_{\rm eff})$ values.
 In the UHF calculations, we use multiple initial conditions, including randomized 
 spin densities, 
 to facilitate convergence to the global minimum~\cite{Xu2011}.

\begin{figure*}[t]
    \centering
    \includegraphics[width = 1\textwidth]{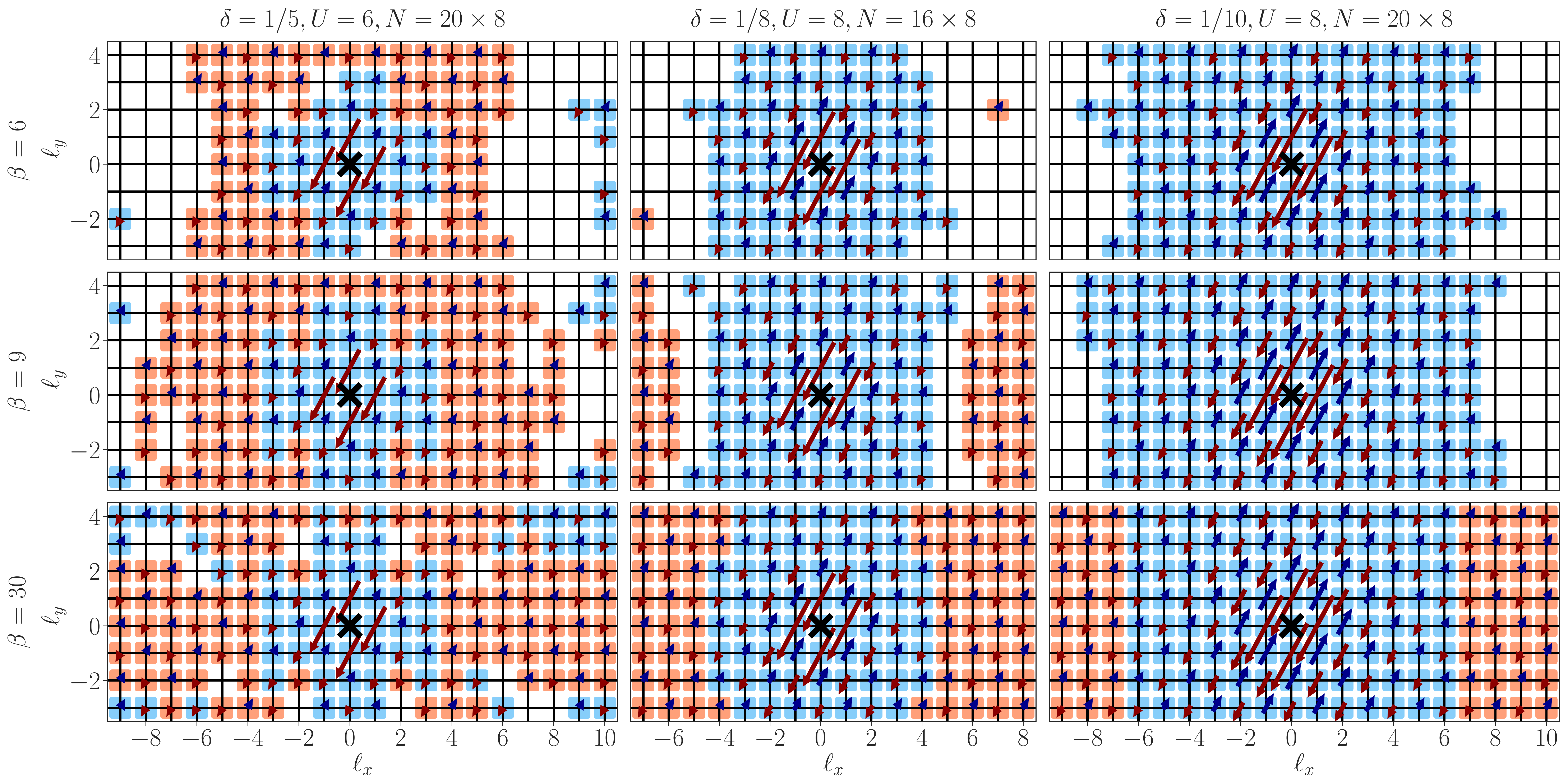}
    \caption{Evolution of the spin correlation function $C_{s}(\ell_{x}, \ell_{y})$ as $T$ is lowered ($T=1/6$, $1/9$, and $1/30$ from top to bottom panels), for doping $\delta = 1/5$, $U = 6$ (left column), $\delta = 1/8$, $U = 8$ (middle), and $\delta = 1/10$, $U = 8$ (right).  These are obtained from our first type of calculations, using fully periodic supercells along both directions, with the correlation averaged over all reference sites ${\bf i}$, indicated by the black cross. (The amplitude of $C_{s}(0, 0)$ is not shown.)  The color of the arrows indicates the sign (dark blue $=+$ and purple $=-$), while the length of the arrows is proportional to the amplitude of $C_{s}(\ell_{x}, \ell_{y})$.  The color of the square at each site represents the sign of the \emph{staggered} spin correlation $(-1)^{\ell_{x} + \ell_{y}}C_{s}(\ell_{x}, \ell_{y})$. A missing arrow and missing colored square indicate that 
    $|C_{s}(\ell_{x}, \ell_{y})|$ is less than twice its Monte Carlo error bar,
    which is considered approximately zero within our statistical resolution.
    }
    \label{fig:fig2}
\end{figure*}

We illustrate the self-consistent scheme in Fig.~\ref{fig:fig1}. 
A $32 \times 4$ lattice is used, with spin pinning fields applied to its left edge, at $\rho = 0.875$ and $U = 8$, targeting an inverse temperature $\beta = 18$.  We start the iteration with the UHF trial density matrix corresponding to the physical parameters.  At $U_{\rm eff} = 8$ and $\rm \beta_{eff} = 18$, the UHF solution at the thermodynamic limit is a diagonal stripe state \cite{Xu2011}.  This can sill be seen in the first step AFQMC result, with some frustration, as shown in the upper color sketch in panel (a).  The UHF solution also greatly over-estimates the strength of the spin order, as seen in the curve with blue circles.  It takes approximately four iterations between AFQMC and IP calculations for the crossover to take place from diagonal to vertical stripes in the trial density matrix, after which the self-consistent process continues to systematically remove the overestimation of spin order in the trial Hamiltonian, until convergence at the solution given by the curve with diamonds (spin pattern illustrated in lower color sketch in panel (a)).  In panel (b), we show the trajectory of the self-consistency iteration for the IP calculations on a heatmap of $\chi^{2}(U_{\rm eff},\beta_{\rm eff})$ as defined in Eq.~(\ref{eqn:self-consistent}).
Such a self-consistency procedure is performed for each system at each temperature, 
in all our calculations of the second type (see Sec.~\ref{ssec:2approaches}).

\section{Results}
\label{sec:results}

In this section we present our main findings, first providing an overview of 
the property of  spin correlations in real space (Sec.~\ref{ssec:spin_correlation}), followed by presenting the picture in momentum 
space with spin susceptibilities and connecting it to real-space signatures (Sec.~\ref{ssec:spin_susceptibility}), across a wide range of temperatures. As mentioned, we systematically investigate three 
doping levels $\delta = 1/5, 1/8$ and $1/10$, i.e., electron densities of $\rho=0.8, 0.875$, and $0.9$, respectively. Then in Sec.~\ref{ssec:short_range_order} we focus on the case of $\delta = 1/5$, 
which is shown to have only short-range order, to examine the temperature evolution of the spin and charge correlations in this physical regime, as well as to illustrate the delicate requirement computationally to determine short versus long-range correlations. Finally in Sec.~\ref{ssec:long_range_order} we discuss
in detail the case of $\delta = 1/8$, to probe the relation between spin and charge orders and the transition temperatures for long-range ordering.

\subsection{\label{ssec:spin_correlation} FINITE-TEMPERATURE SPIN CORRELATIONS IN REAL SPACE}

We first investigate the development of spin-spin correlation as a function of temperature in supercells under fully periodic conditions (i.e., without pinning fields). The supercell size is chosen to depend on $\delta$ to ensure commensuration with the predicted/expected wavelength,
$2/\delta$, of the stripe or SDW state, should such an order exist in the ground state \cite{Zheng2017,Xu2022}. 
In Fig.~\ref{fig:fig2},
we use a $20 \times 8$ supercell for $\delta = 1/5$, which can accommodate two complete SDWs,  
a $16 \times 8$ supercell for $\delta = 1/8$ and a $20 \times 8$ supercell for $\delta = 1/10$, which can accommodate one complete SDW.
We have performed calculations by varying the supercell size and shape,
 some of which are discussed below or presented in the appendices, to 
examine the effect of breaking $C_4$ symmetry and 
 ensure the robustness of the results. 
In all three cases, we see that short-range AFM correlations are immediately present.
As $T$ is lowered, the correlation length increases, and
the magnitude of $ C_{s}({\bm{\ell}}) $ saturates quickly for small separation ${\bm{\ell}}$, and it grows more slowly at larger ${\bm{\ell}}$ towards its low-temperature limit.
A modulation on the spin-spin correlation, 
signaled by the
$\pi$-phase shift in the staggered spin correlation function 
(i.e., parallel spins across the domain boundaries),
is seen almost as soon as the magnitude of the correlation near the domain boundary
exceeds the statistical resolution.

The behaviors at $\delta = 1/8$ and $1/10$, as shown in the middle and right columns of in Fig.~\ref{fig:fig2}, are qualitatively similar. At an inverse temperature of
$\beta=6$, the latter displays a larger AFM correlation length, as further shown in Sec.~\ref{ssec:spin_susceptibility}. This is
consistent with a stronger effective interaction due to smaller doping
($U=8$ in both systems) and a stronger tendency for AFM correlation. 
As we lower the temperature to $\beta = 9$, a clear nodal line develops in $\delta = 1/8$, showing two domains of AFM correlation, while the $\delta = 1/10$ system remains in a single domain. At the very low temperature of $\beta = 30$,
both systems show two well formed domains in the spin correlation, consistent 
with the expected filled stripe state at zero temperature. The
size of the central AF domain in $\delta = 1/8$ fluctuates between $7 \sim 9$ lattice spacings,
consistent with the ground-state stripe state \cite{Zheng2017}. In 
 $\delta = 1/10$, it is somewhat larger than the expected value of $10$ from ground-state results \cite{Xu2022}, likely due to either residual finite-size effects or the use of the simplest, non-interacting uniform 
 trial density matrix as our constraint in these particular calculations. 
 Comparing the last two temperatures, the central domain appears to ``breathe" at $\beta=9$, especially in the 
 case of $\delta = 1/10$, being a slightly larger prior to the full formation of the outer domains at very low temperatures.  We will examine the spin-spin correlation as a function of temperature more quantitatively with $\delta = 1/8$ in Sec.~\ref{ssec:long_range_order}.

Turning to the case of $\delta = 1/5$ in the left column, we see that the short-range spin-spin correlation 
is weaker,  but a clear modulated AFM pattern is visible even at $\beta = 6$, in contrast with the other two cases.
The main reason for this is that the higher hole concentration significantly reduces the modulation 
wavelength, such that it can be comparable to the correlation length 
even at rather high temperatures. 
As we lower the temperature to $\beta = 9$,  the size of the central AFM domain does not change but the adjacent domain grows.  
At $\beta = 30$, the spin correlation shows a strong tendency to restore  $C_4$ symmetry (within
the confines of the rectangular shape of the supercell).  In fact the correlation seems to show, 
rather than a superposition of two  linear waves, more of an isotropic pattern 
which would be compatible with short-range order. 
In order to reliably distinguish 
short- versus long-range orders in the 2-D limit, 
it is important to have 
a more accurate and systematic characterization of the correlation, using
large and wide simulation cells, as we illustrate in the appendix and in Sec.~\ref{ssec:short_range_order}, 
where it is shown that $\delta = 1/5$ indeed exhibits only short-range order and is 
fundamentally different from the cases with smaller doping.

\begin{figure}[t]
    \vspace{-.35in}
    \centering
    \includegraphics[width = .53\textwidth]{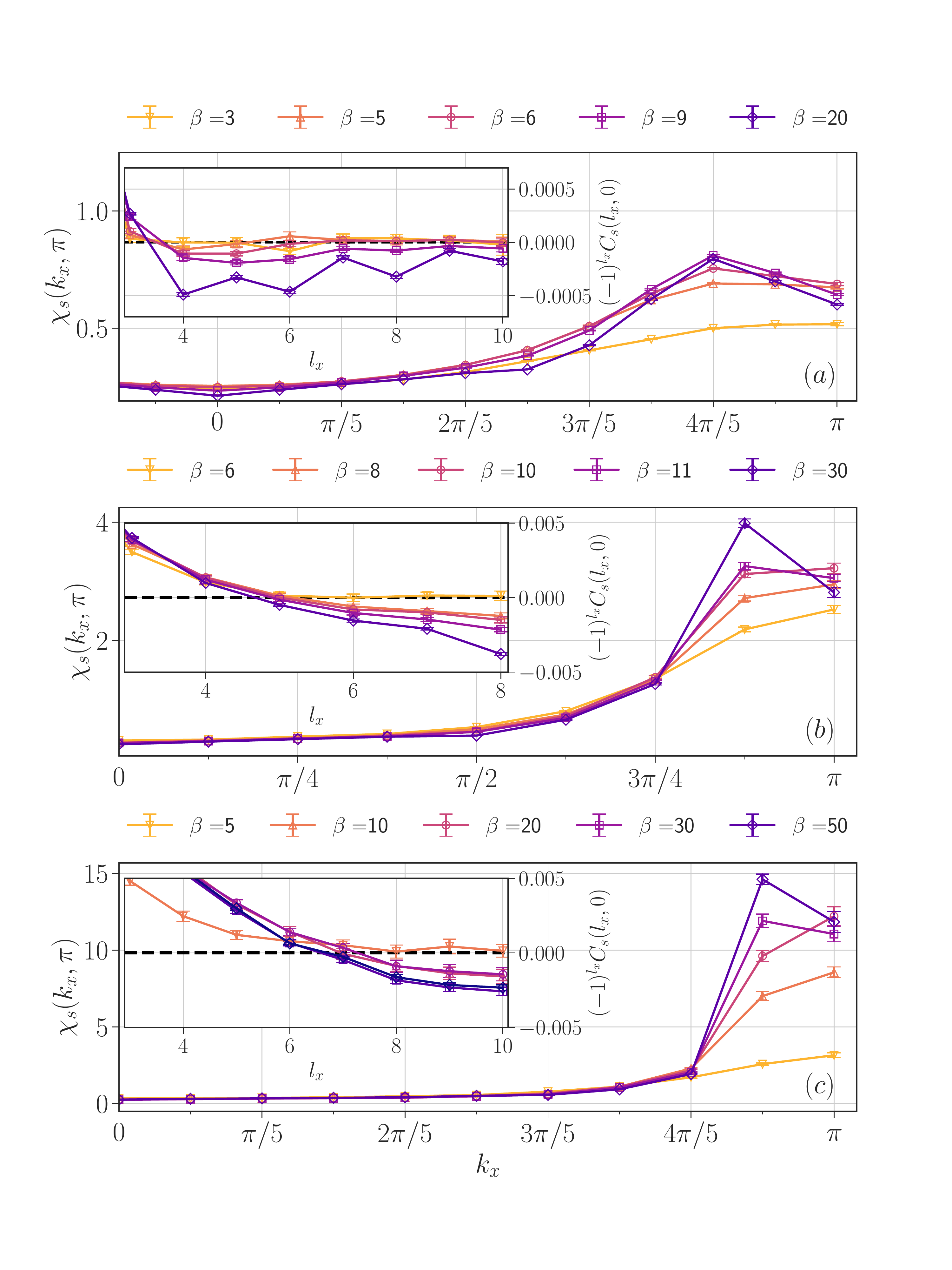}
    \vspace{-.35in}
    \caption{Quantitative characterization of the spin susceptibility $\chi_{s}({\bf k})$ and real-space correlation function $C_{s}(\ell)$ for the three systems in Fig.~\ref{fig:fig2}: a) $\delta = 1/5$, $U = 6$, (b) $\delta = 1/8$, $U = 8$, and (c) $\delta = 1/10$, $U = 8$.  The main figure in each panel shows $\chi_{s}(k_{x}, \pi)$ vs.~momentum $k_{x}$, while the inset shows line cuts of the staggered spin correlation function $(-1)^{\ell_{x}}C_{s}(\ell_{x}, 0)$ along a specific row with $\ell_{y} = 0$. For each system, results for five selected temperatures are displayed.
    }
    \label{fig:fig3}
\end{figure}

\subsection{\label{ssec:spin_susceptibility} SPIN SUSCEPTIBILITY, AND COMPARISON WITH REAL-SPACE SIGNATURES}

The transition from commensurate AF order to incommensurate (IC) SDW order can be characterized by the spin susceptibility $\chi_{s}(k_{x}, k_{y})$ in momentum space, in a complementary manner to the real-space correlation functions.  A shift of the peak value away from $(\pi, \pi)$ 
to $((1-\delta)\pi, \pi)$ and $(\pi, (1-\delta)\pi)$ is a signature of the onset of the SDW. 
However, finite-size effects can affect the detection, either by shifting the incommensurate 
peak position due to size resolution, or removing some of the peaks when rectangular supercells are used. 
Here we analyze the development of spin order as $T$ is lowered using the spin susceptibility $\chi_{s}(k_{x}, k_{y})$, 
and relate and contrast it with the real-space correlation function. 

For  $\delta = 1/8$ both the real-space and momentum-space observables provide a clear signal of the SDW evolution. 
As shown in the inset of Fig.~\ref{fig:fig3}(b), 
the staggered spin correlation function $(-1)^{\ell_{x}}C_{s}(\ell_{x}, 0)$ is zero within error bars for all sites $\ell_{x} \geq 5$ at $\beta = 6$.  When T is lowered to $\beta = 8$, the staggered spin correlation develops a small but clearly resolved signal at larger distances, 
with a change of sign at  $\ell_{x} \sim 5$.  This creates 
a node separating two AFM domains and a $\pi$-phase shift  
across the domain boundaries.   As we continue lowering $T$ down to $\beta = 30$, the magnitude of the correlation function at long distance increases.  However, the position of the node 
only changes slightly, from site-centered to bond-centered.  
From the main graph in Fig.~\ref{fig:fig3}(b), we see that 
the peak position of  $\chi_{s}(k_{x}, k_{y})$ shifts from $(\pi, \pi)$ to $(7\pi / 8, \pi)$
starting from $\beta = 11$.  At $\beta = 30$, the peak at $(7\pi / 8, \pi)$ is dominant, 
consistent with a long-range IC SDW order in the ground state.

The results at $\delta = 1/10$ provide an instructive comparison. 
At $\beta = 5$, the system only displays short-range AFM order, with
larger correlation length than in $\delta = 1/8$, 
as can be seen from the inset
in Fig.~\ref{fig:fig3}(c) and comparison with the inset in panel b. 
In momentum space, the peak values of $\chi_{s}(k_{x}, k_{y})$ are substantially larger than in $\delta = 1/8$, again consistent with stronger AFM correlations.
When we lower the temperature to $\beta = 10$, a node appears between $\ell_{x} = 7$ and $8$ 
in the staggered spin-spin correlation.
The spin susceptibility $\chi_{s}(k_{x}, k_{y})$, 
on the other hand, splits its peak at $(\pi, \pi)$ and 
shifts to $(9\pi / 10, \pi)$ between $\beta = 20$ and $30$.
We argue that the lack of precise correspondence between these different signatures is a feature of such calculations, both because of the correlation being algebraic or weaker, and because of finite-size 
effects which become more severe for lower doping.
At very low temperature $\beta = 30$, the position of the node in the real-space correlation 
shifts back to between $\ell_{x} = 6$ and $7$. 
Note that the spin correlations in real-space at high $T$ could have two subtly 
different behaviors: a modulation is always present even when the correlation is decaying 
exponentially; or it is a purely AFM correlation with no nodes until a crossover occurs to 
an SDW state \cite{Wietek2021}. In momentum space both scenarios would show a signal of the spin susceptibity shifting
from $(\pi, \pi)$ to incommensurate peaks. 
Despite the small but discernible shrinking of the central AFM domain with $T$ mentioned above, 
our results are more consistent with the former physical picture, namely of the modulation appearing as soon as the correlation length is such that a node can be accommodated.

We next contrast these results with the case of $\delta = 1/5$, in  Fig.~\ref{fig:fig3}(a).  The magnitudes of the peak in the spin susceptibility are much lower than for $\delta = 1/8$ and $1/10$, indicating weaker AFM correlations as mentioned earlier. 
At high temperature $\beta \simeq 6$, the peak location starts to shift from $(\pi, \pi)$ to $(4\pi / 5, \pi)$, as nodes in the real-space staggered correlation develop roughly in sync.  These signatures both point to the development of modulated AFM correlation. However, without careful 
finite-size scaling analysis, 
neither can exclude 
the possibility of only short-range order, which turns out to be the 
case at $\delta = 1/5$ and $U=6$, as we show in the next section.
\begin{figure}[h]
    \hspace{-.2in}
    \centering 
    \includegraphics[width = .5\textwidth]{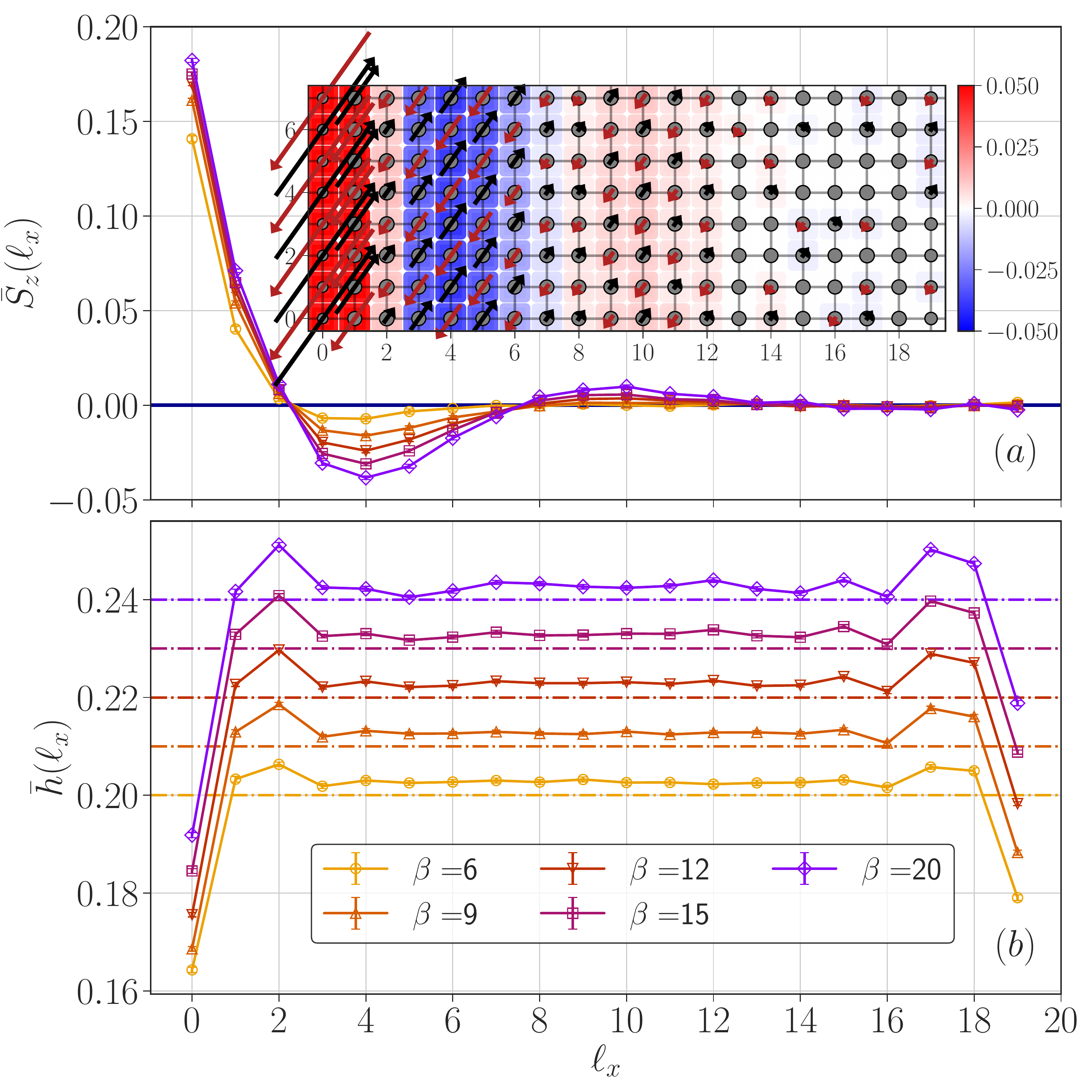}
    \caption{The nature of spin and charge orders with $\delta = 1/5$ and $U = 6$. The results are obtained with our second type of calculations. A spin pinning field is applied to the left edge ($\ell_{x} = 0$) of a $20 \times 8$ supercell. The top panel main graph shows the rung-averaged staggered spin density $\bar{S}_{z}(\ell_{x})$, 
    at five temperatures.  The bottom panel shows the corresponding hole density (note vertical shift for clarity).  The inset in (a) shows an illustration of the spin and hole densities in the supercell at the lowest temperature $T=1/20$.  The color of the arrows indicates the direction, while the length is proportional to the magnitude of $S_{z}(\ell_{x}, \ell_{y})$.  The diameter of black circles is proportional to the hole density $h(\ell_{x}, \ell_{y})$. }
    \label{fig:fig4}
\end{figure}

\subsection{\label{ssec:short_range_order} SHORT-RANGE SPIN AND CHARGE ORDERING IN THE OVERDOPED REGIME} 
 
Compared to lower doping levels,  the $\delta=1/5$ system has the clearest signal of the modulated AFM correlation, which shows up at rather high $T$ and requires  relatively small system size to detect.  At low temperatures the magnitude of the spin correlation function $C_{s}({\bm{\ell}})$ remains substantial up to $|{\bm{\ell}}| \sim 10$. It is then tempting to conclude that this system exhibits prototypical SDW or even stripe order.  However, a more detailed characterization of the spin and charge correlations reveals a different picture.

We apply our second approach, by introducing a local symmetry-breaking field in a cylindrical simulation cell and measuring 
the induced local order parameter.   
The spin and hole densities are used as proxies for the corresponding correlation functions, which magnifies the signal. Furthermore, this approach allows us to employ the self-consistent scheme described in Sec.~\ref{ssec:SC-finiteT}, which provides a better
constraint in AFQMC and leads to more accurate and robust results.
We use an $N = 20 \times 8$ lattice with cylindrical symmetry (OBC in x direction and PBC in y direction). 
The use of OBC in one direction can help probe correlations at larger distances than PBC, if the boundary effect is more localized. 
Spin pinning fields with amplitude $v = 0.1$ are added on the left edge of the cell, 
as described in Sec.~\ref{ssec:2approaches}.
Both the spin pinning fields and boundary condition break symmetry and favor the unidirectional spin- and charge- density waves 
to grow along the cylinder. As mentioned, the combined use of this and 
our first approach, together with finite-size checks varying the shape and size of the 
cell, helps to remove potential bias in the results.
We note that even a width-6 cylindrical cell shows significant finite-size effects which would 
suggest different physics from widths 8 and 10, as illustrated in Appendix~\ref{appendix:width_dependence}, so it is crucial to reach sufficiently large 
simulation cells.

Figure~\ref{fig:fig4} shows how the spin and hole densities evolve as temperature is lowered from $\beta = 6$ to $\beta = 20$. A modulated AFM spin pattern is seen in response to the local spin pinning fields. The magnitude grows with inverse temperature $\beta$ at short range, and displays a non-zero signal at $\ell_{x} \sim 10$, consistent with the 
results from $C_{s}({\bm{\ell}})$ discussed earlier.
At long range, the magnitude of the spin pattern saturates quickly with $\beta$, and the rung-averaged staggered spin density $\bar{S}_{z}(\ell_{x})$ remains zero within statistical errors. As we will see in the 
next section, this is in sharp contrast with the behavior in $\delta=1/8$, where a long-range stripe order develops at $T=0$.

In the charge channel, when a long-range charge stripe exists, holes accumulate on the edges of AFM domains, which leads to peaks in hole density at the nodal points of the staggered spin density. In Fig.~\ref{fig:fig4} (b), we observe a peak at $\ell_{x} = 2$ which corresponds to the first node of the modulated AFM spin pattern.  However, no evidence is seen for further peaks in the hole density at other nodal positions.  (A peak also appears at the opposite edge of the cylinder, which is induced by the OBC. We have performed systematic checks with separate calculations using longer cylinders to gauge and remove its effect.) This indicates that the charge order at $\delta = 1/5$ is extremely short-ranged.  We conclude that the spin and charge correlations remain short-ranged at $\delta = 1/5$, $U=6$ at all temperatures, consistent with the phase diagram from ground-state AFQMC calculations \cite{Xu2022}, an inhomogeneous DMFT study \cite{Peters2014} and the recent higher-$T$ CDet diagrammatic Monte Carlo results~\cite{Simkovic2021}.  It should be emphasized that our results are for $U/t = 6$, to allow direct comparisons with other work \cite{Mai2022}. Increasing the interaction strength will eventually drive the short-range spin and charge orders to become long-ranged in the ground state, with $U/t \sim 8$ at the border separating the two regimes ~\cite{Xu2022}.

\subsection{\label{ssec:long_range_order} INTERPLAY BETWEEN LONG-RANGE SPIN AND CHARGE ORDERING}

\begin{figure*}[t]
    \centering
    \includegraphics[width = 1 \textwidth]{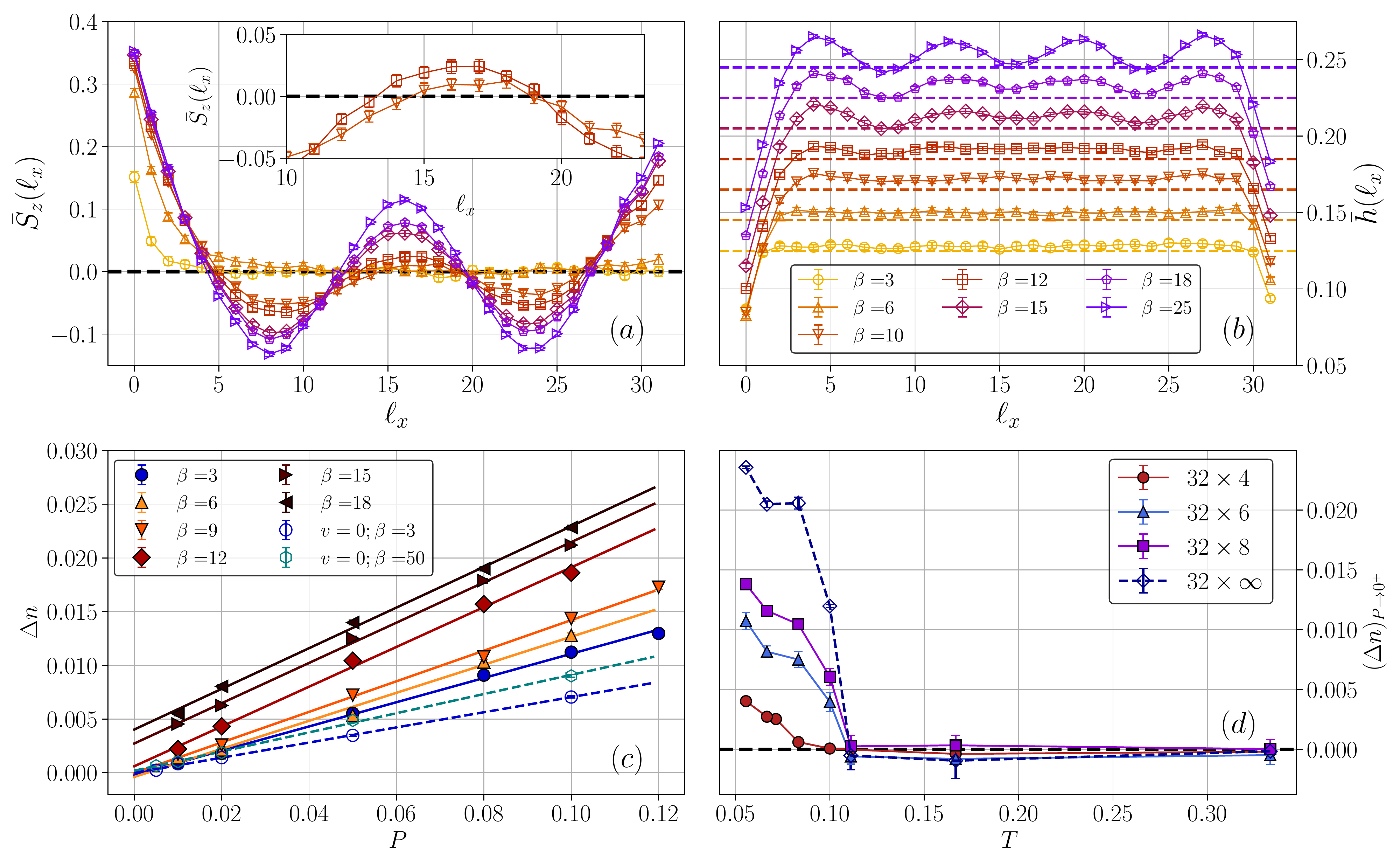}
    \caption{The nature of spin and charge orders with $\delta = 1/8$ and $U = 8$. The top two panels have similar layout as in Fig.~\ref{fig:fig4}. The inset in (a) shows a zoom in the middle region with $\beta = 10$ (triangle) and $\beta= 12$ (square), when the first nodes of spin modulation develop.
    Panel (c) shows calculations to probe the response of the system to an applied external periodic charge potential with amplitude $P$. Solid lines are obtained in the presence of an additional finite edge pinning field for spin. In contrast, dashed lines show otherwise identical calculations, but without the spin pinning field.  Panel (d) shows the development of charge order as  $T$ is lowered, based on the extrapolation procedure shown by the solid lines in panel (c).  Results in panels (a), (b) and (c) are for a $32 \times 4$ supercell, while panel (d) shows three different sizes, $32 \times 4$, $32 \times 6$, and $32 \times 8$ as labelled, as well as an extrapolation to infinite transverse size.}
    \label{fig:fig5} 
\end{figure*}

We now investigate systematically the spin and charge orders, and their interplay, in the case of $\delta = 1/8$ at $U = 8$, applying our second approach of symmetry-breaking pinning fields. 
%
In order
to gauge finite-size effects and deduce the properties at the thermodynamic limit, we
perform large-scale calculations on a sequence of supercell sizes, $N = 32 \times4$, $ 32 \times 6$, and $32 \times 8$,
reaching very low temperatures.
As shown in Fig.~\ref{fig:fig5}(a), at high temperatures ($\beta = 3$ and $\beta = 6$), the staggered spin density displays short-range antiferromagnetic correlations  without an SDW modulation.  When $T$ is lowered, the correlation length grows along with the size of the AFM domain.  Correspondingly, in Fig.~\ref{fig:fig5}(b), the distribution of hole density is uniform at these temperatures except at the ends of the lattice, adjacent to the pinning fields or to the boundary.

When $T$ is lowered to $\beta = 10$ and $\beta = 12$, multiple AFM domains form and the staggered spin density displays a $\pi$-phase shift at domain wall boundaries.  The SDW wavelength is $2/\delta = 16$ and that of the CDW is $1/\delta = 8$.  At intermediate temperature, the SDW modulation starts from both open ends and extends to the central part of the simulation cell.  As shown in the inset of Fig.~\ref{fig:fig5}(a), as $T$ is lowered, the spin correlation length grows and the SDW modulation develops as soon as the correlation length becomes larger than the size of an single AFM domain.  This is consistent with the observation from spin-spin correlation functions, as discussed in Sec.~\ref{ssec:spin_susceptibility}.

A distinguishable AFM domain in the center starts forming at $\beta = 12$, but has a size smaller than the expected periodicity (half a wavelength).  Meanwhile, the wave-like modulation in the charge channel is very weak.  Peaks in the hole density can only be observed at the spin nodal positions closest to the two ends ($\simeq \ell_{x} = 4$ and $\ell_{x} = 27$), and have very small amplitudes.  At $\beta = 25$, two complete incommensurate spin density waves form with the expected stripe periodicity.  Furthermore, a modulation in hole density appears which leads to a charge density wave, with the hole density peak locations coinciding precisely with the spin nodes.  The results at $\beta = 25$ are fully consistent with the ground state properties in this system \cite{Xu2022}, and provide a smooth ``handshake" to $T=0$.

Charge order is seen to follow spin in their temperature evolution in the calculations shown in panels (a) and (b). There is no charge order before the appearance of the first spin node at $\beta\sim 10$. However, the signature of the charge order, namely the oscillation of the hole densities, is very small at intermediate temperatures where
the transition takes place, which makes it challenging to extract more quantitative information. 
In order to further investigate the interplay between spin and charge orders,
we next apply a periodic charge potential as a perturbation in a periodic supercell. 
We perform two kinds of calculations: (i) using this perturbation in the presence of 
the spin pinning fields on one edge, i,e., applying $H_s+H_c$; (ii) using only $H_c$ 
without the finite local spin pinning field.

The results of these calculations are summarized in Fig.~\ref{fig:fig5}(c) and (d).  We define the strength of the induced charge-density-wave as $(\Delta n)_{P} = 1/2 \{n_{\max} - n_{\min}\}$, with $n_{\max} = 1/3 \sum_{\Delta \ell_{x} = -1, 0, +1} \bar{n}(\ell_{x} = \ell_{x}^{\rm max}  + \Delta \ell_{x})$ and $n_{\min} = 1/3 \sum_{\Delta \ell_{x} = -1, 0, +1} \bar{n}(\ell_{x} = \ell_{x}^{\rm min} + \Delta \ell_{x})$, where $\ell_{x}^{\rm max}$ and $\ell_{x}^{\rm min}$ are an expected location for the maximum and minimum density, respectively, according to the minimum and maximum of the applied potential in $H_c$ (at $\ell_{x}^{\max} = 8, 16, 24$ and $\ell_{x}^{\min} = 12, 20$ in the Fig.~\ref{fig:fig5}(b)).  We exclude sites that are near the edges, which eliminates the effect from the spin pinning field in calculations of (i) (while having no effect in calculations of (ii)).  Calculations are performed for a sequence of field strengths, $P$, to obtain $(\Delta n)_P$. The results are shown in Fig.~\ref{fig:fig5}(c) for a set of 8 temperatures using (i), and 2 temperatures using (ii). Extrapolations are performed to the limit of vanishing amplitude of the periodic potential, $P \rightarrow 0$, using a linear fit.
In (i), we find that the resulting $(\Delta n)_{P\rightarrow 0}$ is approximately zero above $T/t \approx 0.1$, 
but becomes non-zero and increases rapidly at lower $T$.  On the other hand, in (ii) when there is no spin pinning,
$(\Delta n)_{P\rightarrow 0}$ remains zero at a temperature as low as $\beta = 50$.
(Note that at high temperature ($\beta=3$), the presence of spin pinning has little effect, and
the two sets of calculations yield similar results, given by the blue lines.) 
These calculations allow us to disentangle the charge order from spin order. The results 
show, rather unambiguously, that the charge order is driven by the spin correlations.

We next examine the transition temperature of the charge ordering.  Fig.~\ref{fig:fig5}(d) plots the extrapolated charge order $(\Delta n)_{P\rightarrow 0}$ from calculations in (i), as a function of temperature, for three supercell sizes.  We vary the width of the supercell from $4$ to $8$, while keeping the length fixed at $32$, which is sufficient to accommodate four (two) full waves of the charge (spin) order in the ground state \cite{Xu2022}. 
The extrapolated charge response remains zero at high $T$ for all system sizes,  but turns finite abruptly at $T \sim 1/10$.  (In contrast, we have verified that the corresponding calculation for spin, applying a periodic spin perturbation for inducing an SDW in a $64\times 4$ periodic supercell, leads to no extrapolated spin order down to $T=1/20$; see Appendix~\ref{appendix:periodic spin pinning}.)

The response at low $T$ becomes stronger as the system size is increased.  The results from the three sizes fit well a linear form in $1/N$ (i.e., $1/4$, $1/6$, and $1/8$), and we indicate the estimate for $L_y\rightarrow \infty$ from this fit by the dashed line in Fig.~\ref{fig:fig5}(d) (See Appendix ~\ref{appendix:width_dependence} for more details on the extrapolation).  These results suggest a finite-temperature phase transition into a phase with long-range (or quasi long-range) charge stripe ordering, with critical temperature $T_{c} \sim 1/10$.  A more systematic finite-size scaling analysis, however, is required in order to determine $T_c$ more accurately, which is extremely challenging given the long-wavelength nature of these orders.  The precise nature of how the spin and charge correlations evolve from high temperature to the ground-state long-range order is subtle.  Our results show that the spin order is the driving force, yet a finite $T_c$ is seen for a transition to charge ordering.  We comment on the possible nature of this transition in the next section.

\section{\label{sec:conclusion} DISCUSSION AND CONCLUSION}

Determining the properties of the doped 2D Hubbard model is faced with enormous challenges, 
that continue to motivate the development of ever more powerful and accurate computational methods. 
At zero temperature, intertwined competing orders are only separated by tiny energy scales, 
of the order of $10^{-2} t$ or less, which can now be resolved by advanced wave-function based methods~\cite{Zheng2017}. 
At finite temperature, progress has been achieved using a variety of methods 
such as cluster extensions of DMFT\cite{Maier2005,tremblay2006,kotliar2006rmp,Gull2010,wu2018,Mai2022}, 
diagrammatic Monte Carlo~\cite{wu2017,Simkovic2021}, DQMC\cite{Huang2017,Huang2018},
as well as METTS~\cite{Wietek2021}.  Nonetheless, limitations in the range of temperature that can be accessed and/or in the system sizes that can be studied have prevented a comprehensive physical picture of the crossovers or transitions in the spin and charge correlations down to $T=0$.

In this work, we have overcome these limitations to realize a full `handshake' between finite-$T$ and ground-state calculations, by deploying the latest advances in the AFQMC approach.  This includes in particular the development of a self-consistent constraint with an effective finite-temperature mean-field performed by using both an effective interaction $U_{\rm eff}$ and an effective inverse temperature $\beta_{\rm eff}$.  This new approach allows us to reach supercell sizes and temperatures which were previously inaccessible, namely lattices of many hundred sites and temperatures as low as $1/50$.  The ability to perform computations which can span the full range of temperatures and connect with ground-state methods adds an important dimension to the accurate treatment of strongly correlated systems, and will be useful in many other systems besides the Hubbard model.

Systematic and meticulous checks were carried out by testing different forms of trial density matrices, using supercells with different geometries and boundary conditions and performing calculations both for translationally invariant systems and under symmetry-breaking external fields.  As a result, we have obtained an accurate and systematic description of the physical nature and $T$-dependence of the spin and charge correlations over an extended range of temperature at three representative values of doping and interaction strength: $\delta = 1/5$ at $U/t=6$ and $\delta=1/8$, $\delta=1/10$ at $U/t=8$.

We showed that, at the largest doping $\delta =1/5$ and $U/t=6$, SDW correlations start to develop at rather high $T$, but remain short-ranged as $T$ is lowered.  Correspondingly, no charge ordering is found except for very short-range correlations.  These results are consistent with the very recent findings of Ref.~\cite{Simkovic2021}, but much lower $T$ was reached in the present study.  At the two smaller doping levels and $U/t=8$, spin correlations first develop at the antiferromagnetic wave-vector $(\pi,\pi)$ at high-$T$ and an incommensurate SDW is found with lowering temperature, as soon as the correlation length is sufficient to accommodate a node.  These results confirm  the qualitative expectations from mean-field theory~\cite{Zaanen1989, Poilblanc1989, Machida1989, Schulz1990, Kato1990} and are broadly consistent with recent computational results~\cite{Wietek2021,Simkovic2021}.  We find that, at low temperatures the spin and charge correlation have modulation wavelengths compatible with the ground-state results at these parameters, i.e., filled stripes.  In each case our results extend to very low temperatures and accomplish a full handshake with $T=0$ properties.

Crucially, our work establishes how this growth of SDW correlations eventually connects to the ground-state stripe order at $T=0$ \cite{Zheng2017, Xu2022}.  We identify a finite-temperature phase transition below which charge ordering sets in.  This transition takes place at a fairly low temperature $T_c/t< 0.1$ and was therefore inaccessible to previous studies.  It will be valuable to confirm our result with more systematic finite-size scaling studies, which will require further algorithmic advances.  By different calculations that impose  bulk periodic potentials and/or boundary pinning fields, we were able to probe the different charge and spin responses to study their interplay. We establish that charge order is driven by the development of spin stripe correlations.

The nature of the critical behavior at the charge ordering transition poses questions of broad theoretical interest.  This is extremely challenging to resolve numerically, because of the enormous length scale needed.  However our results do place some bounds on possible scenarios.  At a given commensurate doping $\delta=1/q$, there is a discrete symmetry $\mathbb{Z}_q$ corresponding to the translations of the stripe pattern with wavelength $q$, e.g. $\mathbb{Z}_8$ for $1/8$-doping.  It is interesting to note that the critical behaviour of the two-dimensional $\mathbb{Z}_q$-symmetric $q$-state clock model has a non-trivial dependence on $q$.  While for $q<5$ the phase transition corresponds to the breaking of the discrete $\mathbb{Z}_q$ symmetry with long-range order for $T<T_c$, an emerging $U(1)$ symmetry has been suggested to occur for $q\geq 5$, leading to a low-temperature phase with algebraic quasi long-range order and a BKT transition~\cite{ortiz}.  Hence, one could imagine two different scenarios. The most straightforward one is that long-range charge order takes place below $T_c$, while the spin correlation length remains finite at any non-zero $T$ and long-range spin ordering takes place only at $T=0$, in accordance with the Mermin-Wagner theorem \cite{ Mermin1966, Hohenberg1967}. 
 Another more intriguing scenario is that $T_c$ corresponds to a BKT transition below which the charge has algebraic quasi long-range order, and that the coupling between the charge and the spin degrees of freedom also leads to algebraically decaying spin correlations in the low-$T$ phase.  These questions are left open for a future study and in all likelihood will require and trigger significant new developments in computational methods.  Another obviously crucial and still rather open question is the interplay between the charge and spin orders discussed here and superconductivity.  For a mean-field Landau theory of the interplay between charge and spin ordering, see Ref.~\cite{Zachar1998}.

From an experimental standpoint, our work has implications in several directions.
Experiments with ultracold fermionic atoms in optical lattices combined with site-resolved microscopy are 
able to image spin and charge correlations at progressively lower temperatures in a setup which directly emulates the Hubbard model
with nearest-neighbor hopping only, as studied here~\cite{Bohrdt2021,Koepsell2019,Hartke2022}. 
Indeed, in these experiments, growing antiferromagnetic correlations as temperature is lowered have been demonstrated 
in both the half-filled and doped systems~\cite{Mazurenko2017}, consistent with our finding that 
the onset of spin {\it correlations} takes place at a higher temperature than for charge correlations. 
Our determination of the temperature below which charge order occurs provides guidance for future experiments, 
which 
may be able to address this issue soon 
thanks to progress in cooling techniques.

Our work also has connections to experimental results on the cuprate materials. 
Indeed, for cuprates that display long-range stripe {\it ordering} such as Nd-LSCO~\cite{Ichikawa2000}
and LBCO~\cite{Abbamonte2005}, spin long-range order 
develops at a temperature below that of the charge ordering~\cite{Tranquada2020} in agreement with our finding.
Note however that the stripes seen in our results are `filled', while in La-based cuprates, 
they are found to be `half-filled'~\cite{Lee2022}. 
This difference, as well as other qualitative differences, points to the importance of the next-nearest neighbor 
hopping $t^\prime$. Extending our work to include this important parameter, as well as considering 
more realistic models such as a three-band Hubbard model including oxygen states are extremely valuable directions in 
which to extend the work presented here.

\acknowledgements

We would like to thank Richard Scalettar and Fedor {\v S}imkovic for carefully reading the manuscript and insightful suggestions. We are also grateful to Michel Ferrero, Edwin Huang, Andrew Millis, Subir Sachdev and Hao Xu for very useful discussions. Y. Y. H. acknowledges
the National Natural Science Foundation of China
(NSFC) under Grant No. 12047502. The Flatiron Institute is a division of the Simons Foundation. 

\appendix
\section{\label{appendix:trial_density_matrix} Evolution of trial density matrix in the self-consistent scheme}

\begin{figure*}
    \centering
    \includegraphics[width = .95\textwidth]{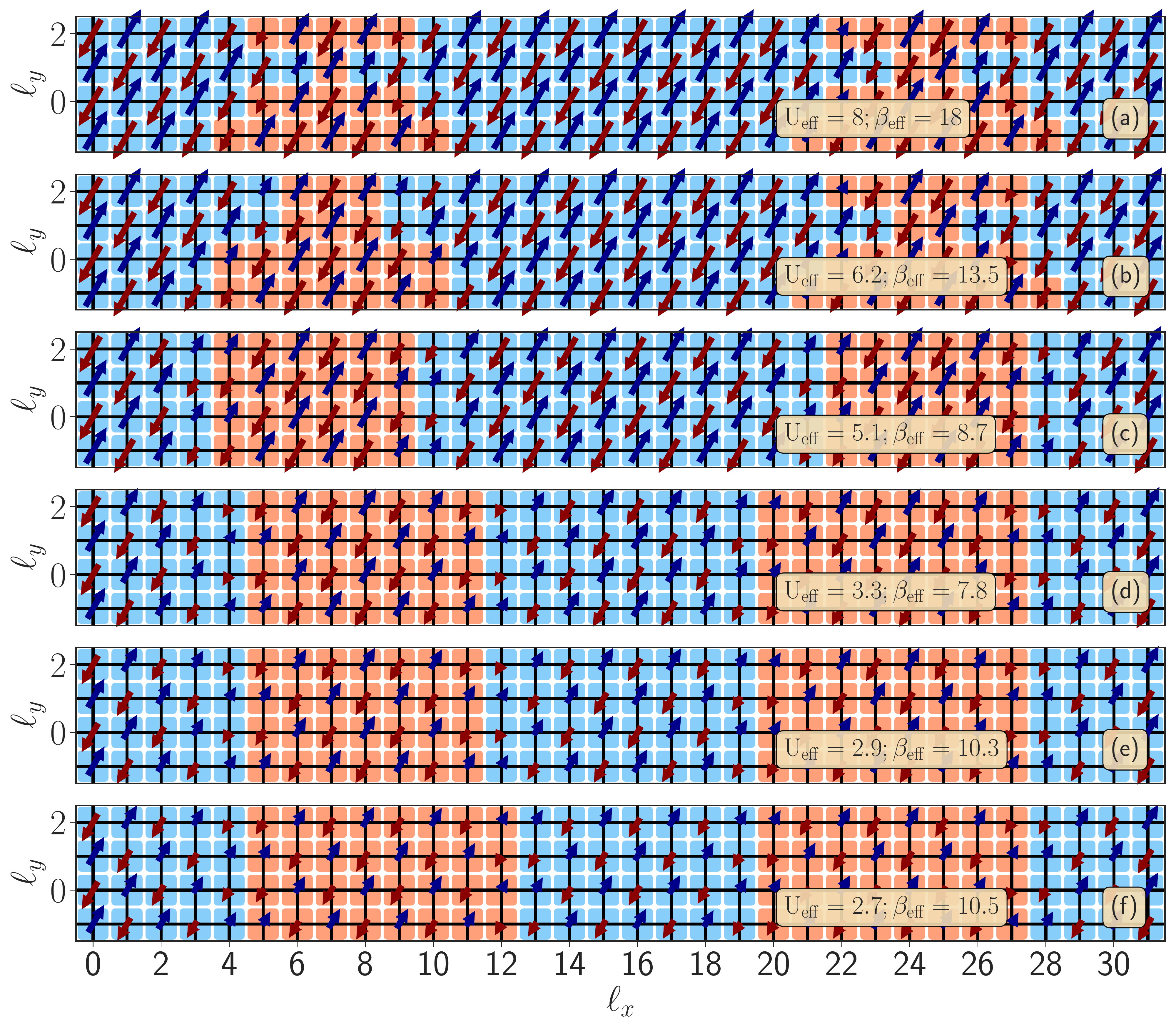}
    \caption{(a)-(f) Convergence of the trial state in the self-consistent process which couples it to the AFQMC calculation.
    The spin density in real space is shown for the UHF solution at selected $(\rm U_{eff}, \beta_{eff})$ values 
     on the optimization path shown in Fig.~\ref{fig:fig1}.  Each UHF solution is used to form the input trial density matrix $H_{T}$ for the next step AFQMC calculation in the iteration. The trial state starts from an incorrect diagonal stripe order and converges to the correct order from the self-consistent coupling to AFQMC.
     The system is $32 \times 4$ with $\delta = 1/8$ and the amplitude of edge pinning fields is $v = 0.1$.
    }
    \label{fig:fig6}
\end{figure*}

In the main text, we have shown the optimization path on the $\rm U_{eff}$-$\rm \beta_{eff}$ plane of the UHF solutions when we perform the self-consistent scheme which couples the many-body calculation (AFQMC) with an effective single-particle calculation (UHF). 
Here we provide additional details and illustrate how the input trial density matrix $H_{T}$ evolves and converges to its final configuration.
In Fig.~\ref{fig:fig6}, we show this in real space with an 
example of a 
$32 \times 4$ simulation cell,  
at fixed electron density $\delta = 1/8$, with AF pinning fields on the left edge.

We start the self-consistent scheme by using the 
physical parameters in the single-particle calculation, 
$\rm (U_{eff}, \beta_{eff}) = (U, \beta) = (8, 18)$, 
where $U$ and $\beta$ are the parameters in the many-body 
calculation.  As shown in Fig.~\ref{fig:fig6}(a), the self-consistent solution of UHF at this point is a diagonal SDW modified by the pinning fields, which is consistent with previous results \cite{Xu2011, Inui1991}.  Using this UHF configuration as the initial input trial density matrix, we obtain spin-resolved electron density on each site through AFQMC calculation. We determine the $\rm (U_{eff}, \beta_{eff})$ for the next iteration by searching on a gird and choosing the point where the difference of spin density between the AFQMC output and UHF solution is minimal.  As shown in Fig.~\ref{fig:fig6}(b), a UHF solution with $\rm (U_{eff}, \beta_{eff}) = (6.2, 13.5)$ is used as the new trial density matrix $H_{T}$, where $\chi^{2}$ is the minimal.  As we continue the self-consistent procedure, the trial density matrix evolves from a diagonal stripe to a vertical stripe at $\rm (U_{eff}, \beta_{eff}) = (5.1, 8.7)$, as shown in Fig. \ref{fig:fig6} (c).  However, the size of the antiphase domains fluctuates and is not equal to a half of the expected SDW wavelength of 8.
As we continue this procedure, the magnitude of the spin densities decreases 
and the SDW wavelength saturates to the correct value with small fluctuations.  The self-consistent procedure ends when $\rm (U_{eff}, \beta_{eff})$ no longer changes with the statistical resolution, which indicates that a fixed point is reached.  The final (single-particle) trial density matrix is a vertical stripe which has very similar shape to the many-body solution.

\section{\label{appendix:square_lattice_geometry} The effect of C$_4$ rotational symmetry}

\begin{figure}
    \centering
    \includegraphics[width = .5\textwidth]{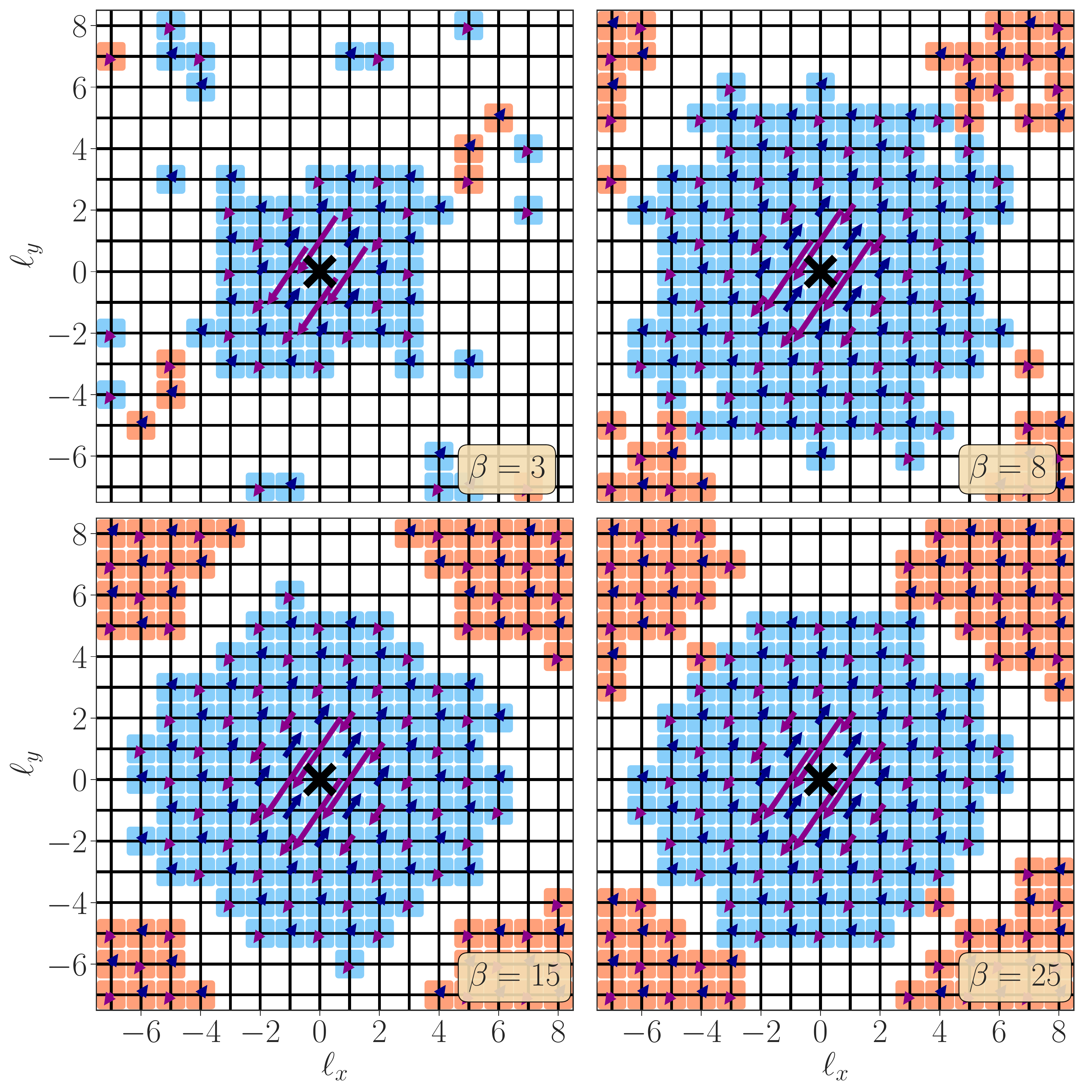}
    \caption{Evolution of the spin correlation function $C_{s}(\ell_{x}, \ell_{y})$ as $T$ is lowered.  The results are for a $16 \times 16$ square lattice with $U = 8$ and $\delta = 1 / 8$ doping, at four $T$ values as labeled (in $\beta$). 
}
    \label{fig:fig7}
\end{figure}

In addition to the cylindrical cells,
we also performed calculations in the square geometry to cross-check the results. 
In Fig.~\ref{fig:fig7}, we show an example in a $16 \times 16$ supercell with PBC in both directions.  The evolution of the spin correlation function is depicted
as $T$ is lowered.  At high temperature $\beta = 3$, only a single AFM domain can be observed in the center of the square lattice.  The reference point is labeled by a black cross.  As $T$ is lowered to $\beta = 8$, the size of central AFM domain increases significantly and antiphase regions appear at the corners, which is quantitatively consistent with the results obtained on rectangular lattices as shown in Fig.~\ref{fig:fig2}.  In the square supercell, the rotational symmetry is preserved, and the pattern of the spin correlation function is a superposition of a horizontal stripe and a vertical one.  As $T$ is further lowered to $\beta = 15$ and $\beta = 25$, the central AFM domain becomes a sharp diamond and the antiphase regions in the corners become larger and saturate.

\begin{figure}
    \centering
    \includegraphics[width = .5\textwidth]{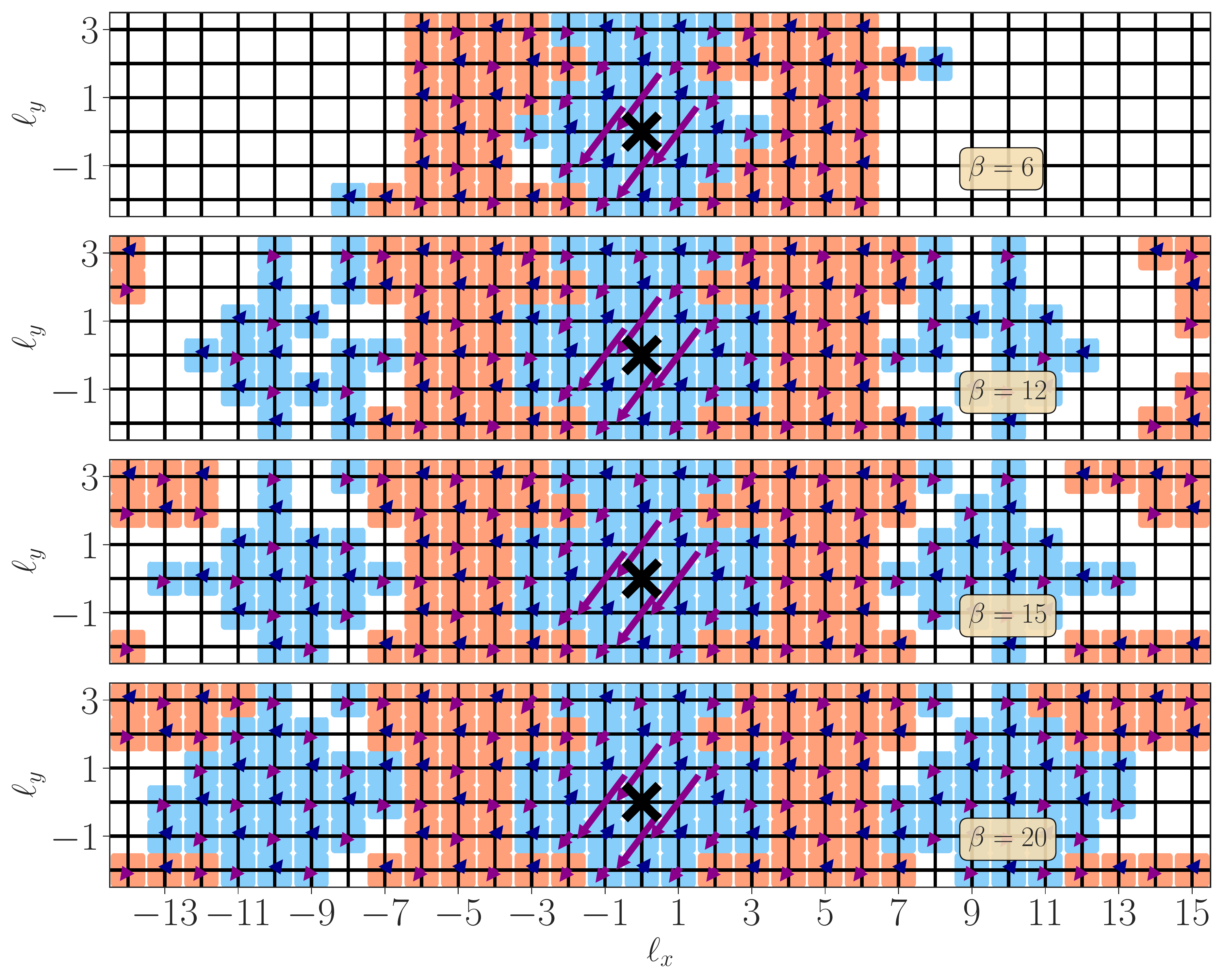}
    \caption{Evolution of the spin correlation function $C_{s}(\ell_{x}, \ell_{y})$ as $T$ is lowered, at $\delta = 1 / 5$, on a $30 \times 6$ lattice with $U = 6$.  PBC is used in both directions.  AFM regions are observed, and the results in this finite system do not distinguish short- vs.~long-range order.
    }
    \label{fig:fig8}
\end{figure}


\section{\label{appendix:width_dependence}  Study of finite-size effects and approach to the thermodynamic limit} 

As discussed in the main text, finite-size effects can crucially alter the nature of the correlation and give misleading information about the property in the thermodynamic limit (TDL). For example, in many situations, a width-4 cylinder can give qualitatively different physics \cite{Chung-plaquette,Xu2011}.
Width-6 cylinders are often much better; however in the case of 
$\delta = 1 / 5$ it turns out that even a width-6 cylinder is insufficient for predicting the correct order in the TDL.

\begin{figure}
    \centering
    \includegraphics[width = .5\textwidth]{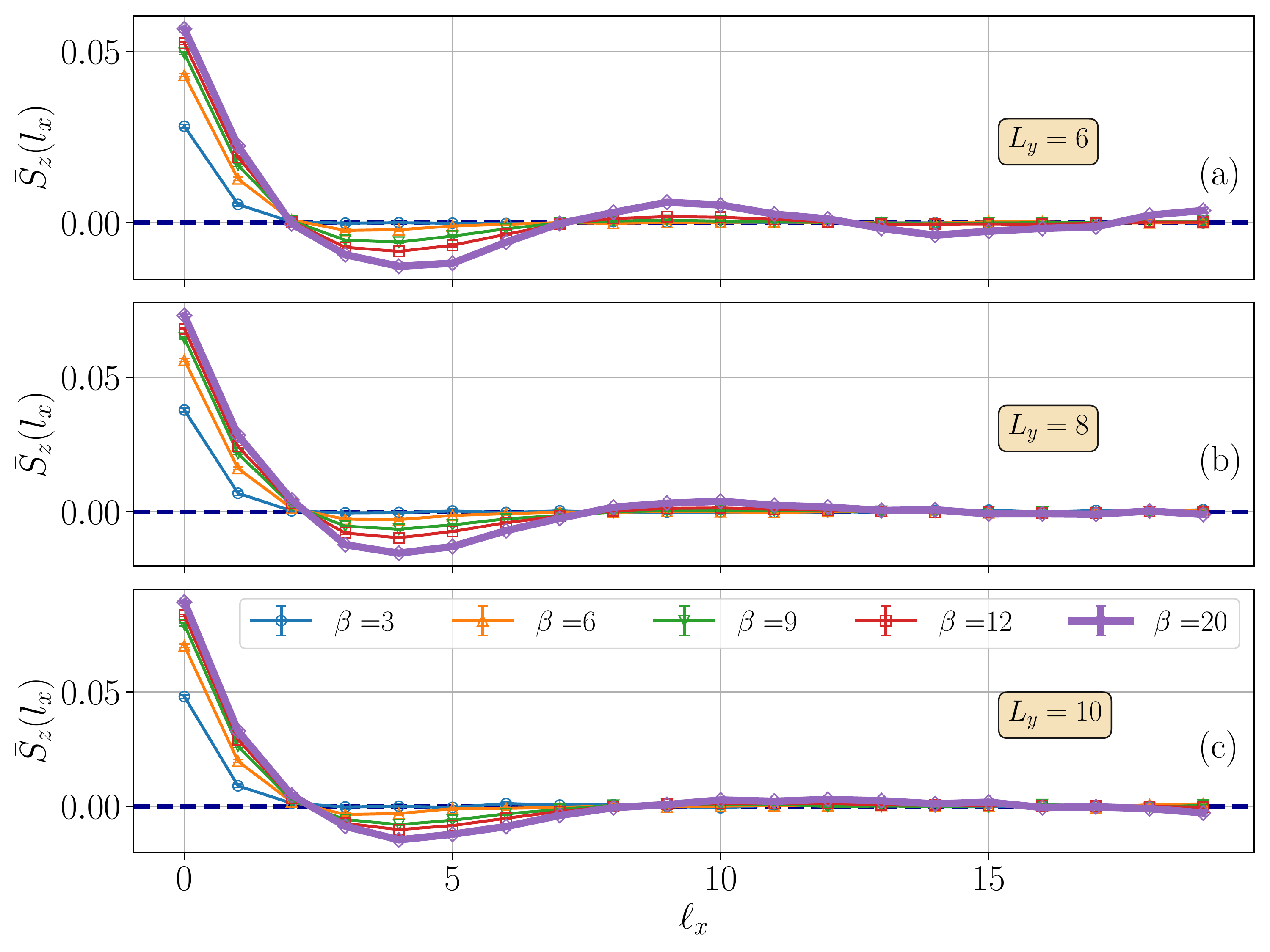}
    \caption{Investigating the behavior in the TDL at $\delta = 1 / 5$ ($U = 6$). 
    Rung-averaged staggered spin density $S_{z}(\ell_{x})$ is plotted vs.~site $\ell_{x}$ on cylinders with fixed length $L_{x} = 20$ and varying width.  Staggered pinning fields with amplitude $v = 0.1$ are applied to the left edge, at $(\ell_{x} = 0)$.  The three panels show widths (a) $L_{y} = 6$, (b) $L_{y}$ = 8, and (c) $L_{y} = 10$, respectively.  Converged results obtained from the self-consistent procedure are plotted at each temperature.  Error bars are smaller than the symbols size.}
    \label{fig:fig9}
\end{figure}

\begin{figure*}[t]
    \centering
    \includegraphics[width = 1\textwidth]{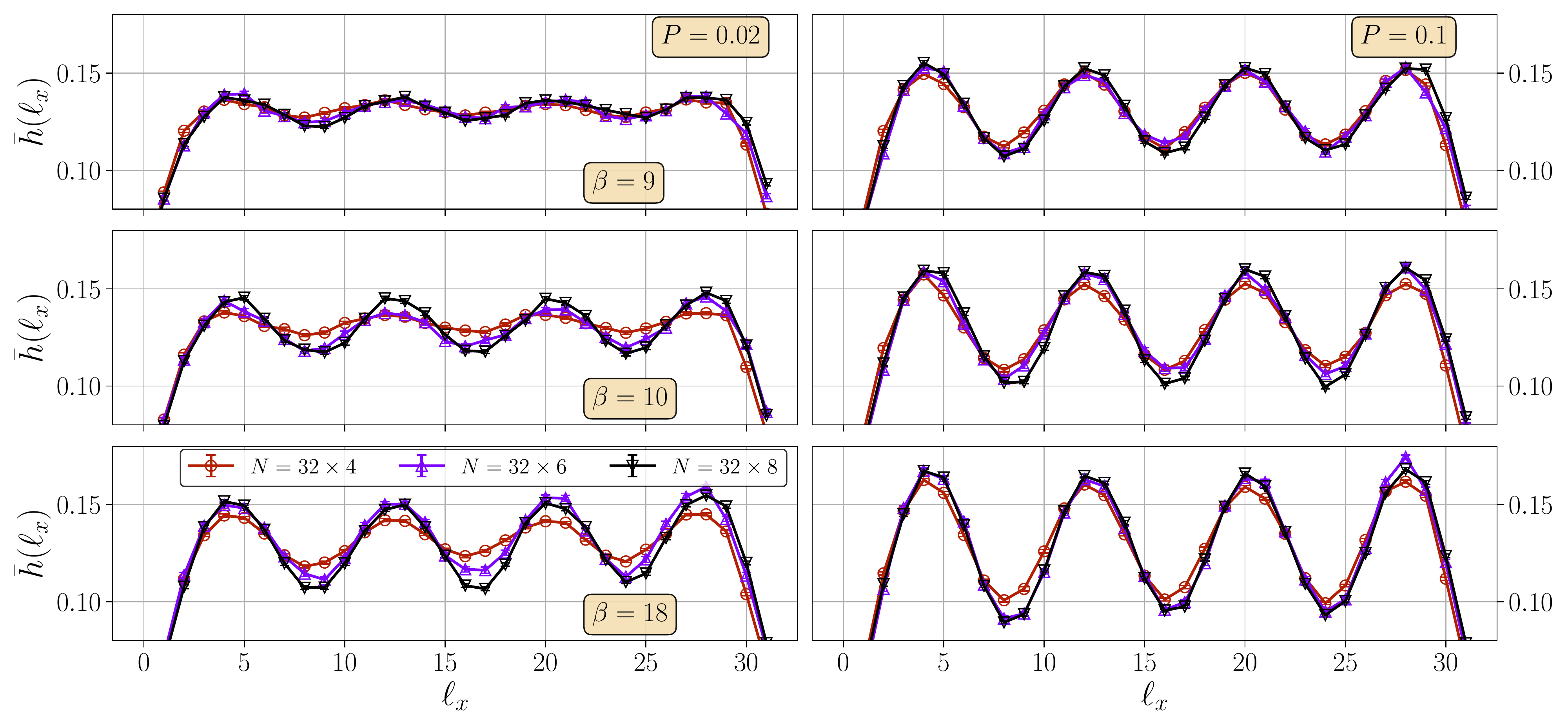}
    \caption{
    Finite-size analysis of the charge response to an external periodic charge perturbation. Rung-averaged hole densities $\bar{h}(\ell_{x})$ are shown for two strengths ($P$) of the external charge potential, at three temperatures spanning the transition $T_c$ (see discussions in Fig.~\ref{fig:fig5}).  The systems are at $\delta = 1 / 8$ and $U = 8$, with $L_x=32$ and staggered spin pinning fields of amplitude $v = 0.1$ applied to the left edge at $\ell_{x} = 0$.  Three lattice sizes are studied, with width $L_{y} = 4, 6$ and $8$.  Error bars are smaller than symbol size. 
    }
    \label{fig:fig10}
\end{figure*}

In Fig.~\ref{fig:fig8}, we show the temperature evolution of spin order at  $\delta = 1/5$, $U = 6$ on a $30 \times 6$ lattice.  In contrast to $\delta = 1 / 8$, antiphase regions 
form even at temperatures as high as $\beta = 6$.  The width of each AFM region 
fluctuates from $\Delta \ell_{x} = 3$ to $\Delta \ell_{x} = 7$ sites in different rows.  As $T$ is lowered to $\beta = 12$ and $15$, more AFM regions appear but the amplitude of spin correlation at long distance is small.  At $T = 0.05$, the sizes of the third AFM regions (symmetric about the center) tends to saturate to that of the central one.  These behaviors are similar to the results in Ref.~\cite{Mai2022}, in which a $16 \times 4$ cluster is used in DCA, and would suggest that there is no apparent distinguishing features between them and those in $\delta=1/8$ or $1/10$ (except that the signature for AFM modulation or stripe appears more easily, at higher $T$).

However, a different picture emerges as we increase the width of the simulation cells to reach the TDL.  In Fig.~\ref{fig:fig9}, we use supercells with fixed length $L_{x} = 20$ but different widths $L_{y} = 6, 8$ and $10$.  To determine the spin order more accurately, we add AFM spin pinning fields to the simulation cell and measure the one-point function (the rung-averaged spin density).  For each temperature, we employ the self-consistent scheme and plot the final converged results.  On the width-$6$ cylinder, the result is compatible with that of the correlation function shown in Fig.~\ref{fig:fig8}.

In sharp contrast, when we increase the width of the cylinder to $L_{y} = 8, 10$, we find that the spin order is short-ranged.  The high-temperature behavior of the spin density on width-$8$ and width-$10$ cylinders is the same as that on the width-$6$ cylinder.  As $T$ is lowered, the spin density at large distance $(\ell_{x} \geq 8)$ is essentially zero with error bars as shown in Fig.~\ref{fig:fig9} (b) and (c).  This is consistent with the conclusion from ground-state calculations \cite{Xu2022}.

In Fig.~\ref{fig:fig10}, we show the size dependence of the hole density in response to an external periodic 
perturbation which couples to the density.  Two representative amplitudes of the periodic charge potential are shown, $P = 0.02$ and $P = 0.1$.  We select two temperatures ($\beta = 9$ and $10$) near the estimated phase transition and one temperature ($\beta = 18$) deep in the charge ordered phase, based on the analysis in Fig.~\ref{fig:fig5}.  
At higher temperature ($\beta = 9$), where charge order is absent, we see a negligible size dependence, as the 
charge response is essentially an independent-electron response to the external potential.  As we lower $T$ and approach the transition, the finite-size dependence becomes different for different perturbation strengths, showing a larger correction at $P = 0.02$. This trend becomes even stronger at $T=1/18$ (well below the transition), when the size dependence remains small at $P = 0.1$ while a large increase is seen in the amplitude of the induced CDW with increasing supercell size at  $P = 0.02$.


\begin{figure}
    \centering
    \includegraphics[width = .5\textwidth]{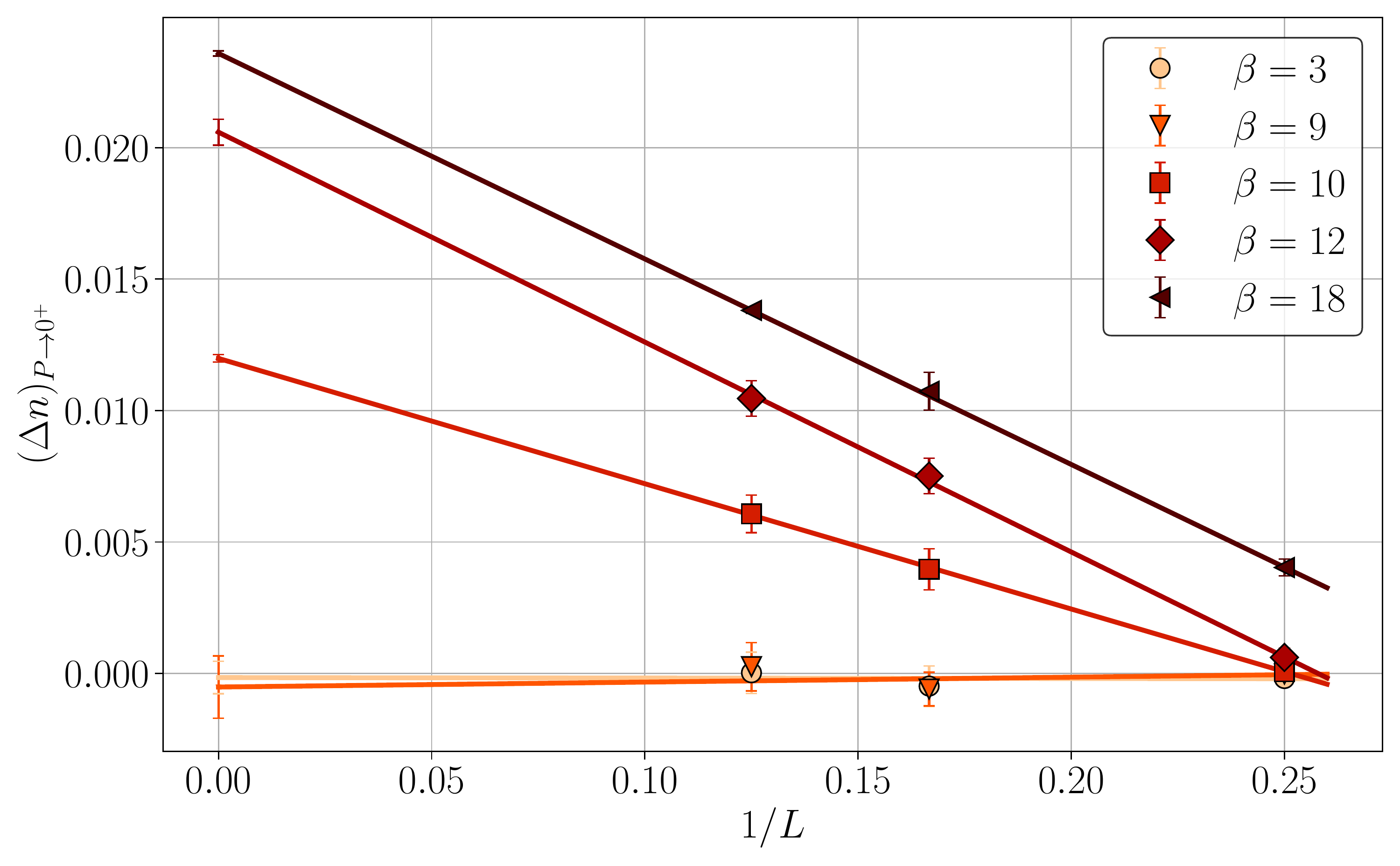}
    \caption{Finite-size extrapolation of the charge order in Fig.~\ref{fig:fig5} (d).  The charge order $(\Delta n)_{P \rightarrow 0^{+}}$ is plotted vs. the inverse of transverse lattice size $1 / L_{y}$ for three different system sizes $N = L_{x} \times L_{y}$ with fixed $L_{x} = 32$ while varying $L_{y} = 4, 6$ and $8$, respectively, at five representative temperatures.  The results of the linear extrapolation to $1 / L_{y} \rightarrow 0$ are shown in Fig.~\ref{fig:fig5} (d).}
    \label{fig:fig11}
\end{figure}

In Fig.~\ref{fig:fig11}, we show the finite-size extrapolation of the charge order (amplitude of the charge response to an external charge perturbation), under the effect of a local spin pinning field at the edge of the supercell.  As mentioned, the dimension $L_x$ of the supercell is fixed at $L_x=32$.  The finite-size data at each $T$ fit very well a linear function of $1/L_y$.  The response at low $T$ becomes stronger as the width of the lattice increases.  The behavior of the response as a function of lattice size 
is consistent with a transition to long-range charge order at a finite temperature.  A more systematic finite-size scaling analysis to determine $T_c$ precisely would require further knowledge of the system and additional computations in significantly larger system sizes.

\section{\label{appendix:structure_factor} Spin and charge structure factors under the two different approaches}

\begin{figure*}
    \centering
    \includegraphics[width = .95\textwidth]{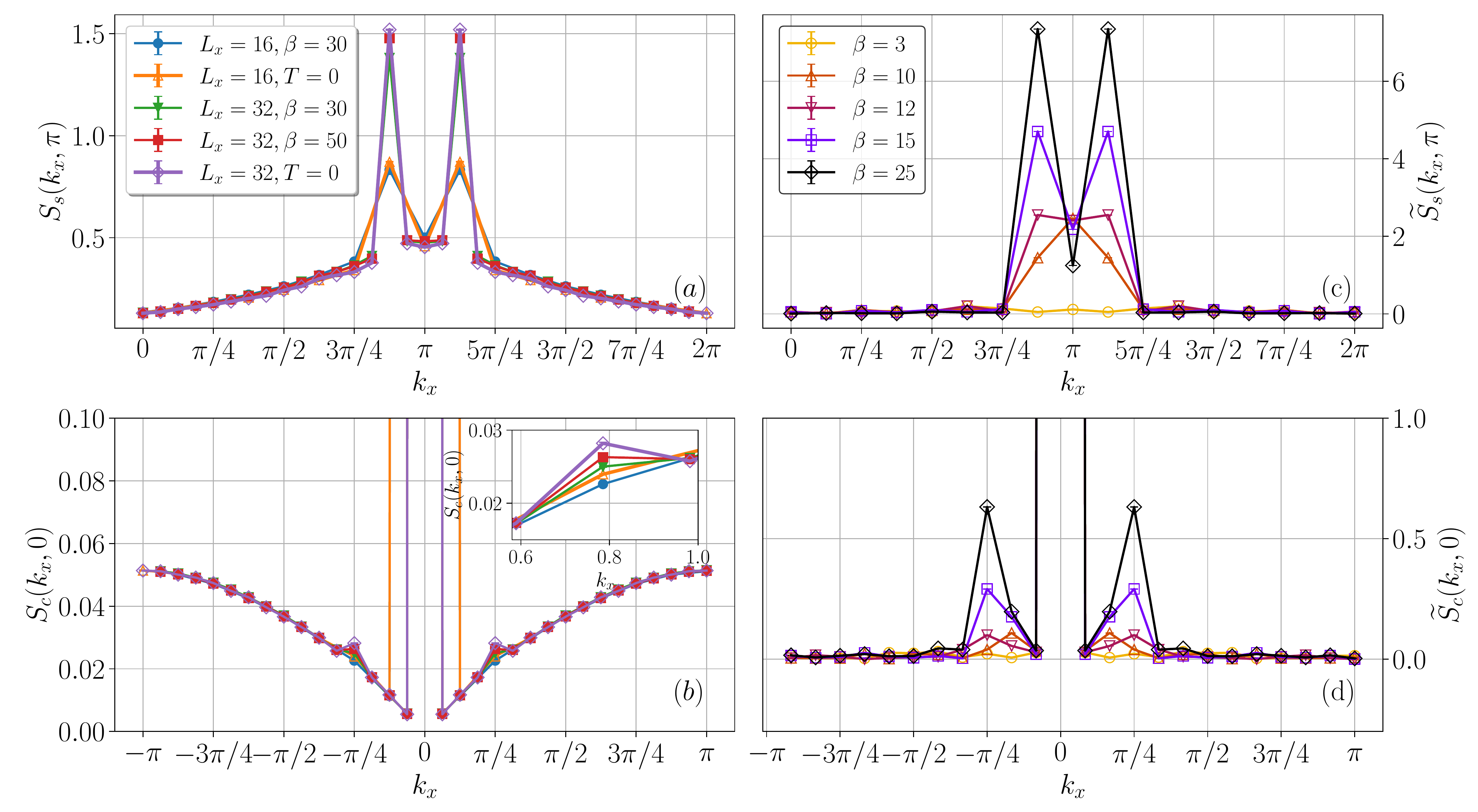}
    \caption{Spin and charge structure factors under the two different approaches, fully PBC  versus pinning field calculations.
    The left column shows (a) spin structure factor $S_{s}(k_{x}, \pi)$ and (b) charge structure factor $S_{c}(k_{x}, 0)$ vs.~momentum $k_{x}$ on $16 \times 4$ and $32 \times 4$ lattices with PBC applied in both directions, for two finite (low) temperatures, compared to  the corresponding ground-state result. 
    The right column shows the Fourier transforms of the (c) spin density $\tilde{S}_{s}(k_{x}, \pi)$ and (d) charge density $\tilde{S}_{c}(k_{x}, 0)$ vs. $k_{x}$ on segments of a $32 \times 4$ lattice, for a sequence of temperatures. 
     $U = 8$ and $\delta = 1 / 8$ are used in all calculations.
    }  
    \label{fig:fig12}
\end{figure*}

We have discussed in the main text the two different approaches to probe spin and charge correlations, namely fully PBC calculations versus pinning field calculations. We have illustrated how they complement each other. It is reassuring that they lead to consistent results on the spin and charge order, even with very different settings and under very different constraining density matrices. However, as we discussed, they can have different behaviors in finite-size systems and give different signal strengths. In this appendix, we show some comparison of the structure factors to supplement the discussions in the main text.

The equal-time structure factors for the spin and charge correlations discussed in the main text are defined by:
\begin{eqnarray}
    \begin{aligned}
        S_{s}({\bf k}) &= \frac{1}{N} \sum_{\ell, m = 1}^{N} 
            e^{i {\bf k} \cdot ({\bf r}_{\bm{\ell}} - {\bf r}_{{\bf m}})} 
            \langle S_{z}(\bm{\ell}) S_{z}({\bf m}) \rangle,  \\
        S_{c}({\bf k}) &= \frac{1}{N} \sum_{\ell, m = 1}^{N} 
            e^{i {\bf k} \cdot ({\bf r}_{\bm{\ell}} - {\bf r}_{{\bf m}})} 
            \langle n_{\bm{\ell}} n_{{\bf m}} \rangle.
    \end{aligned}
\end{eqnarray}
In Fig.~\ref{fig:fig12} (a), we present the spin  structure factor at $U = 8$ and $\delta = 1/8$.  Lattices with two different sizes ($16 \times 4$ and $32 \times 4$) are used.  For both lattice sizes, the peak position of $S_{s}(k_{x}, \pi)$ is seen to shift from $(\pi, \pi)$ to $(7\pi / 8, \pi)$ and $(9 \pi / 8, \pi)$.  
Similar to the spin susceptibility $\chi_{s}$ in Sec.~\ref{ssec:spin_susceptibility}, it indicates 
the evolution of an incommensurate SDW order. 

 The signature for charge order is subtle and more challenging to detect,
as shown in Fig.~\ref{fig:fig12} (b).  On a $16 \times 4$ lattice, 
the amplitude of charge structure factor peak at $(\pm \pi / 4, 0)$ remains extremely small, down to $T = 0$. This reflects strong finite-size effects, with the signal for long-range correlation being 
submerged in a large background. 
To detect the peak that characterizes the charge stripe, we need to use larger lattices, e.g $32 \times 4$.  Then distinguishable but small peaks at $(\pm \pi / 4, 0)$ appear at $T = 0.033$. 
As $T$ is lowered to $0.02$, the peaks become more clear and eventually converge to the amplitude at ground state ($T  = 0$).  

As a comparison, the second approach of applying a pinning 
field to turn the spin and charge density into a proxy for 
correlation functions shows much clearer signals. 
We apply Fourier transform to the one-point functions, 
\begin{eqnarray}
    \begin{aligned}
        \tilde{S}_{s}({\bf k}) &= \sum_{\ell = 1}^{N^{\prime}} e^{i{\bf k}\cdot{\bf r}_{\bm{\ell}}} \langle S_{z}(\bm{\ell}) \rangle,  \\
        \tilde{S}_{c}({\bf k}) &= \sum_{\ell = 1}^{N^{\prime}} e^{i{\bf k}\cdot{\bf r}_{\bm{\ell}}} \langle n(\bm{\ell}) \rangle\,,
    \end{aligned}
\end{eqnarray}
where the spin and charge densities are shown in the main 
text, for example in Fig.~\ref{fig:fig5} (a) and (b).
We perform the Fourier transform on a subset of the lattice, 
with sizes of $16 \times 4$ and $24 \times 4$, respectively, 
for the spin and charge densities. 
Sites on both ends are discarded to remove effects induced by spin pinning fields and the OBC.

As shown in Fig.~\ref{fig:fig12} (c), as $T$ is lowered, the peak of $\tilde{S}_{s}({\bf k})$ gradually shifts from $(\pi, \pi)$ to $(7\pi/8, \pi)$ and $(9\pi/8, \pi)$, which 
signals the development of the SDW order.  In contrast to the charge structure factor, the shoulder peaks in $\tilde{S}_{c}({\bf k})$ at $(\pm \pi/4, 0)$ appear much more clearly, starting approximately at $\beta = 15$, consistent with the result from directly applying charge-inducing field discussed in the main text. Its height is much larger ($\geq 10$ times) than that from 
the charge structure factor. 

\section{\label{appendix:pairing susceptibility} $d$-wave pairing susceptibility at $\delta = 1/5$ and $U=6$}
\begin{figure}[h]
    \centering
    \includegraphics[width = 0.5\textwidth]{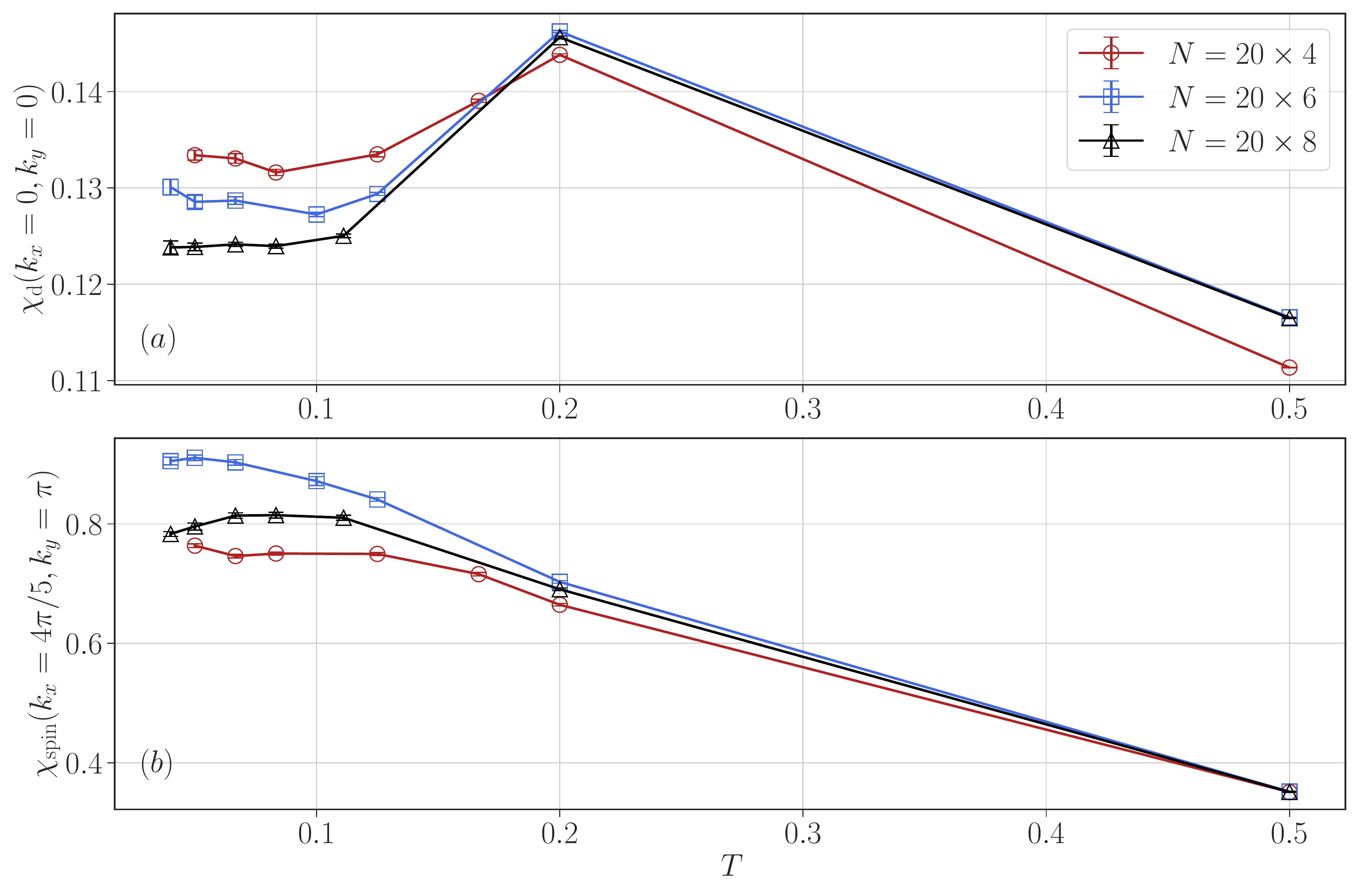}
    \caption{(a) The d-wave pairing susceptibility $\chi_{\rm d}({\bf k})$ plotted at ${\bf k} 
    = (0, 0)$, as a function of temperature $T$ for three different supercell sizes.  
    (b) The corresponding spin susceptibility $\chi_{\rm spin}({\bf k})$ plotted at the characteristic wave-vector for incommensurate SDW. 
    The system has $U/t = 6$ and $\delta = 1/5$.}
    \label{fig:fig13}
\end{figure}

As mentioned, recent work (see, e.g., Ref.~\cite{Qin2021}) indicates that there is no $d$-wave superconducting long-range order in the ground state of the first two systems that we study, at 
$\delta = 1/8$ or $\delta = 1/10$ ($U/t = 8$). The third system, 
$\delta = 1/5$ with $U/t = 6$, is near the boundary of the parameter regime explicitly scanned in Ref.~\cite{Qin2021}); 
indeed there has been a recent suggestion that there may be superconducting order in the ground state \cite{Sorella2021}. 
We have computed in this system 
the $d$-wave pairing susceptibility $\chi_{\rm d}$, which is defined as 
\begin{eqnarray}
    \begin{aligned}
        \chi_{\rm d}({\bf k = 0}) &= \sum_{\ell = 1}^{N} \int_{0}^{\beta} \langle \Delta_{\rm d; {\bf r + r_{\ell}}}^{}(\tau) \Delta_{\rm d; {\bf r}}^{\dag}(0) \rangle d\tau,  \\
        \Delta_{\rm d; {\bf r}}^{\dag} (\tau) &= e^{\tau H} \Delta_{\rm d; {\bf r}}^{\dag}(0) e^{-\tau H},  \\
        \Delta_{\rm d; {\bf r}}^{\dag} &= \frac{1}{4} \sum_{\delta} (-1)^{\delta} c_{{\bf r}, \uparrow}^{\dag}c_{{\bf r}+\delta, \downarrow}^{\dag}  \\
        &= \frac{1}{4} c_{{\bf r}, \uparrow}^{\dag} \left(c_{{\bf r} + \hat{x}, \downarrow}^{\dag} - c_{{\bf r} + \hat{y}, \downarrow}^{\dag} + c_{{\bf r} - \hat{x}, \downarrow}^{\dag} - c_{{\bf r} - \hat{y}, \downarrow}^{\dag} \right).
    \end{aligned}
\end{eqnarray}

In Fig.~\ref{fig:fig13}(a), we show the result 
for three system sizes $N = 20 \times 4$, $N = 20 \times 6$ and $N = 20 \times 8$.  As $T$ is lowered, the d-wave pairing susceptibility first grows and then decreases.  At low temperature $T < 0.1$, as the system size increases, the pairing susceptibility decreases, which is incompatible with the development of long-range order.  (This, of course, cannot rule out the possibility of a superconducting order developing below the temperature scale we investigate.)
For reference,  the corresponding spin susceptibility is shown in panel (b).  As $T$ decreases, $\chi_{\rm spin}(4\pi / 5, \pi)$ increases (except for a small downward bend in 
$20 \times 8$). 
Recall that this system does not have long-range spin order as $T\rightarrow 0$, as we showed in the main text.

\section{\label{appendix:periodic spin pinning} Response to periodic spin potential}

To complement our studies of the charge response in Sec.~\ref{ssec:long_range_order}, we apply periodic spin pinning fields and examine 
the amplitude of the induced spin order in the limit of zero external perturbation in the linear response regime.  Similar to what was done for the charge channel, we apply sinusoidal spin pinning fields to a $64 \times 4$ lattice at $U/t = 8$ and $\delta = 1/8$.  PBC is used in both directions of the simulation cell.  We add the following to the original Hamiltonian 
    \begin{eqnarray}
        H_{s} = \sum_{\ell_{x}, \ell_{y}} B \sin(\kappa \cdot \ell_{x} + \phi) S_{z}(\ell_{x}, \ell_{y}),
        \label{eqn:eqnF1}
    \end{eqnarray}
where $B$ is the amplitude and $\kappa$ is the characteristic wave-vector.  
In this system the wavelength of the SDW 
is 
twice that of the charge stripe, 
so a longer cylinder ($64 \times 4$) is used to accommodate 
the same number of SDWs as for 
charge stripes 
in Fig.~\ref{fig:fig5}.

\begin{figure}[h]
    \centering
    \includegraphics[width = .48\textwidth]{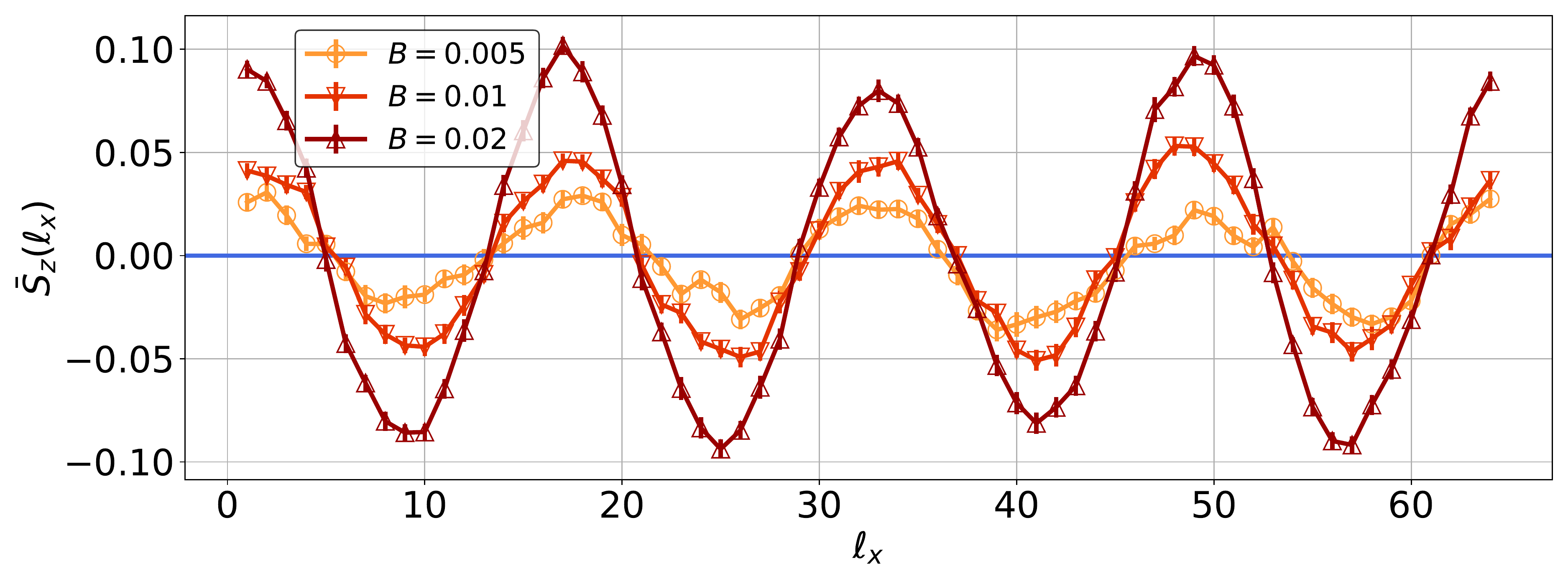}
    \caption{Row-averaged real-space staggered spin density $\bar{S}_{z}(\ell_{x})$ vs. $\ell_{x}$ with different amplitudes of the input spin perturbation.  A $64 \times 4$ lattice is used with $U/t = 8$, $\delta = 1/8$ and $\beta = 20$.}
    \label{fig:fig14}
\end{figure}

In Fig.~\ref{fig:fig14}, we plot the row-averaged staggered spin density $\bar{S}_{z}(\ell_{x})$ as defined in Eq.~(\ref{eqn:eqn6}). 
We observe four complete spin dnesity wave periods, 
with positions of peaks and valleys 
consistent with the applied sinusoidal potential. 
As we increase the amplitude of the input spin perturbation, the amplitude of $\bar{S}_{z}(\ell_{x})$ increases approximately linearly as a response. 
Note that the magnitude of the spin response is much larger than that of the charge, as seen in comparison with Fig.~\ref{fig:fig10}.
To quantify the response 
we define the strength of the induced SDW order as $\Delta S_{z} (B) = 1/2 \left(S_{z}^{\max} - S_{z}^{\min} \right)$, with $S_{z}^{\max} = 1/3 \sum_{\Delta \ell_{x} = -1, 0, +1} \bar{S}_{z}(\ell_{x} = \ell_{x}^{\max} + \Delta \ell_{x})$ and $S_{z}^{\min} = 1/3 \sum_{\Delta \ell_{x} = -1, 0, +1} \bar{S}_{z}(\ell_{x} = \ell_{x}^{\min} + \Delta \ell_{x})$, where $\ell_{x}^{\max}$ and $\ell_{x}^{\min}$ are the expected locations of the maximum and minimum staggered spin $\bar{S}_{z}$, respectively.

\begin{figure}
    \centering 
    \includegraphics[width = .48\textwidth]{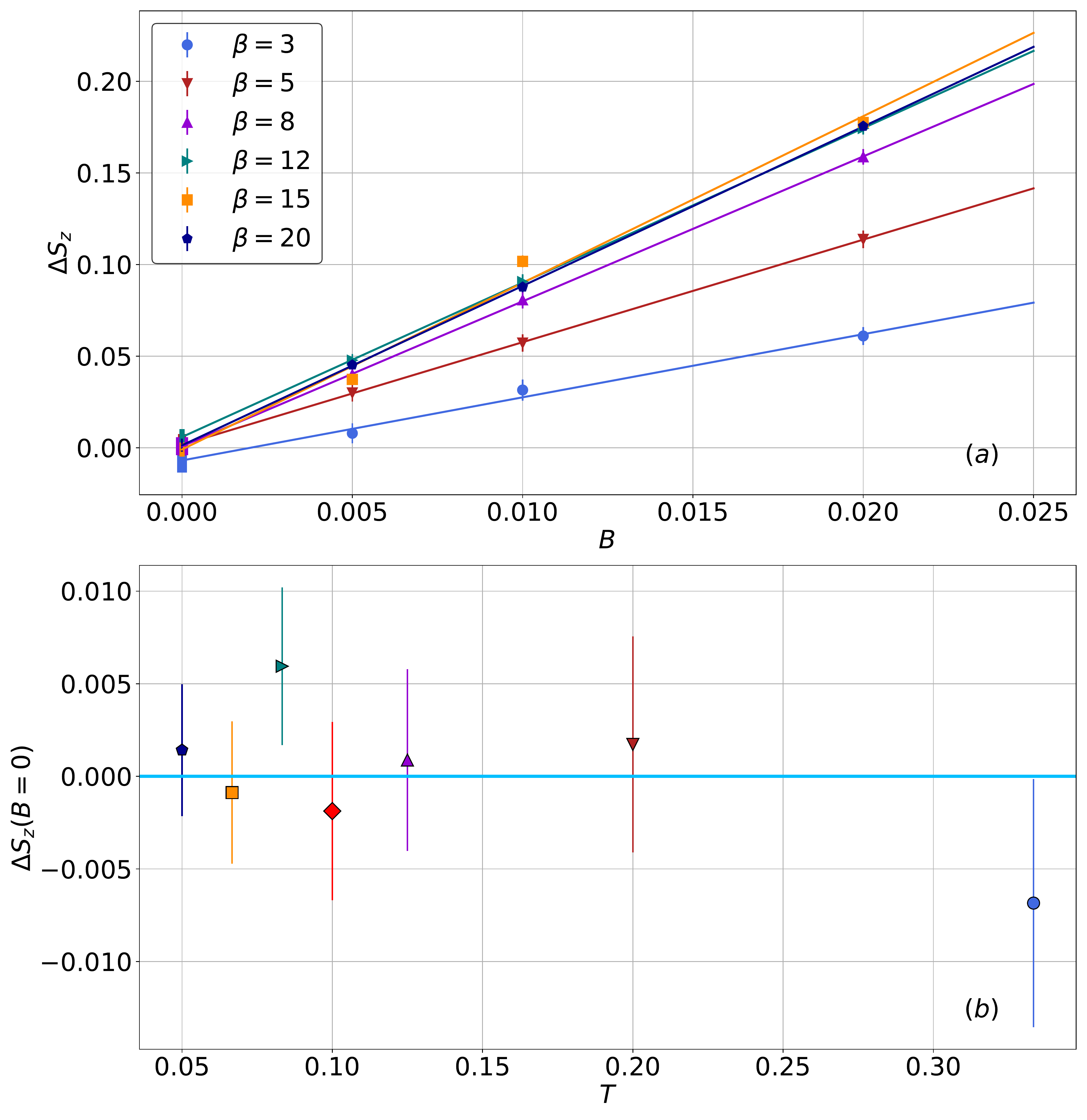}
    \caption{(a) Linear extrapolations for the induced SDW order $\Delta S_{z}(B)$ versus the external potential strength $B$ for six different temperatures, for the same system as in Fig.~\ref{fig:fig14}. 
    (b) The absence of spin order as $T$ is lowered, based on the extrapolation procedure shown in panel (a).  
    }
    \label{fig:fig15}
\end{figure}

As shown in Fig.~\ref{fig:fig15} (a), we then perform extrapolations to 
vanishing amplitude of the periodic spin potential, 
$B \rightarrow 0$, using a linear fit. 
Fig.~\ref{fig:fig15} (b) plots the extrapolated spin order $\Delta S_{z}(B = 0)$ as a function of temperature.
The extrapolated SDW order is 
zero within statistical error bars down to the lowest temperature of $T = 0.05$, 
consistent with Mermin-Wagner theorem. 
The contrasting behaviors between the spin and charge responses provide yet another strong consistency check on our computational approach.

\bibliography{finite_temperature_hubbard}

\end{document}